\pdfoutput=1

\documentclass[preprint]{elsarticle}
\biboptions{sort&compress}
\usepackage[utf8x]{inputenc}
\usepackage{ucs}
\usepackage{amsmath}
\usepackage{amsfonts}
\usepackage{amssymb}
\usepackage{makeidx}
\usepackage{float}
\usepackage{graphicx}
\usepackage{subcaption}
\usepackage{booktabs}
\usepackage{array}
\usepackage{multicol}
\usepackage{changepage}
\usepackage{hyperref} 
\usepackage{booktabs}
\usepackage{lscape}
\usepackage[export]{adjustbox}
\usepackage{blindtext}

\journal{Journal of Computational Physics}

\begin{document}

\begin{frontmatter}
\title{A novel momentum-conserving, mass-momentum consistent method for interfacial flows involving large density contrasts}

\author[1]{Sagar Pal\corref{cor1}}
\ead{sagar.pal@sorbonne-universite.fr}

\author[1]{Daniel Fuster}
\ead{daniel.fuster@sorbonne-universite.fr}

\author[1]{St\'ephane Zaleski}
\ead{stephane.zaleski@sorbonne-universite.fr}

\cortext[cor1]{Corresponding author: Sagar Pal}

\address[1]{Institut Jean le Rond $\partial$'Alembert, Sorbonne Universit\'e and CNRS, Paris, France}

\begin{abstract}
We propose a novel method for the direct numerical simulation of interfacial flows involving large density contrasts, using a Volume-of-Fluid method. We employ the conservative formulation of the incompressible Navier-Stokes equations for immiscible fluids in order to ensure consistency between the discrete transport of mass and momentum in both fluids. This strategy is implemented on a uniform 3D Cartesian grid with a staggered configuration of primitive variables, wherein a geometrical reconstruction based mass advection is carried out on a grid twice as fine as that for the momentum. The implementation is in the spirit of Rudman (1998) \citep{rudman1998volume}, coupled with the extension of the direction-split time integration scheme of Weymouth \& Yue (2010) \citep{wy} to that of conservative momentum transport. The resulting numerical method ensures discrete consistency between the mass and momentum propagation, while simultaneously enforcing conservative numerical transport to arbitrary levels of precision in 3D. We present several quantitative comparisons with benchmarks from existing literature in order to establish the accuracy of the method, and henceforth demonstrate its stability and robustness in the context of a complex turbulent interfacial flow configuration involving a falling raindrop in air.        
\end{abstract}

\begin{keyword}
large density contrasts \sep consistent discrete mass-momentum transport \sep sub-grid advection \sep geometric Volume-of-Fluid \sep staggered Cartesian grids \sep 3D conservative direction-split advection
\end{keyword}

\begin{highlights}
\item Conservative formulation of Navier Stokes with interfaces using the Volume-of-Fluid method.
\item Geometrical interface and flux reconstructions on a twice finer grid enabling discrete consistency
	between mass and momentum on staggered uniform Cartesian grids. 
\item Conservative direction-split time integration of geometric fluxes in 3D, enabling discrete conservation of mass and momentum. 
\item Quantitative comparisons with standard benchmarks for flow configurations involving large density contrasts.  
\item High degree of robustness and stability for complex turbulent interfacial flows, demonstrated using the case of a falling raindrop.   
\end{highlights}

\end{frontmatter}

%
%
%
%
%
%
%
%
%

\section{Introduction}
The dynamics of liquid-gas interfacial flows play a critical role in several processes in nature,
as well as in myriad industrial applications.
The key elements of such flows are droplets and bubbles, that constitute the
fundamental mechanisms governing the exchange of heat and mass at the ocean-atmosphere interface \cite{seinfeld1998air,deike},
mixing/separation in the context of metallurgical processes \cite{johansen1988fluid,metal},
conventional modes of heat transfer \cite{deckwer1980mechanism,bubble}
and ever so importantly, the transmission of pathogens \cite{lydia_1,lydia_2}.

\vspace*{0.2cm}

A substantially large subset of all surface tension dominated flows
involve significant disparities in the material properties across
the interface, the most common example being flow configurations
corresponding to air-water systems, where the densities and viscosities
of the fluids are separated by (approximately) 3 and 2 orders or magnitude, respectively.
The development of numerical methods that attempt to model
such interfacial flows involving marked contrasts in density
face several challenges, key amongst them being the transport of
mathematical discontinuities that arise out of the aforementioned contrasts.
Extremely small numerical errors are ubiquitous as a consequence of the
numerous approximations involved at each and every step of the algorithm
(e.g. interface reconstruction, flux computation etc.) .
In the context of such flows, such ``numerical errors'' may result in physically
inconsistent mass and momentum transfer across the interface, often from
the denser phase towards the lighter phase as a consequence of inadequate numerical resolution.
The presence of large density contrasts tend to amplify
the growth of these cascading numerical errors, eventually
leading to significant (often catastrophic) interfacial deformations,
followed rapidly by a loss of numerical stability.

\vspace*{0.2cm}

Since such numerical instabilities were first observed with
the ``SURFER'' \cite{lafaurie1994modelling} code
in the context of planar jets and rising bubbles,
considerable efforts have been made towards the design of numerical methods
to specifically deal with flows involving such marked density contrasts
(i.e. large $|\log r|$, where $r$ is the ratio of the densities of the two fluids).
The underlying principle behind these endeavours is that the governing
equations for the transport of mass and momentum are solved using a
conservative formulation (divergence of fluxes), instead of standard
non-conservative forms, which themselves were adapted directly from
techniques developed originally for single-phase flows.
This formulation enables one to render the discrete transport of
momentum \textit{consistent} with respect to the discrete transport of mass.
Such a tight coupling of the propagation of errors between the discrete
mass and momentum fields enables alleviation of many of the issues
that plague such numerical methods, especially in the
context of low to moderate spatial resolutions.

\vspace*{0.2cm}

Exacerbating the already complicated nature of the discrete transport of material
discontinuities is the role of surface tension on the evolution of the interface.
They are commonly modeled as singular source terms in the momentum balance equation that
governs the evolution of the velocity field, with the capillary force itself
being proportional to the third derivative of the interfacial position.
Thus, a secondary but important mitigating factor is the advancements
made in the modeling of capillary forces, resulting in the adoption of
consistent and \textit{well-balanced} surface tension formulations.
Consistency in the context of surface tension models refers to the ability
of methods to progressively achieve more accurate estimations of interfacial
curvature as a result of increasing spatial resolution,
whereas well-balanced refers to the ability to recover certain static
equilibrium solutions pertaining to surface tension dominated flows
without the perpetual presence of parasitic or spurious currents in the velocity fields.
We refer the reader to influential works of Popinet \cite{popinet2018numerical,popinet2009accurate}
to get a better understanding of the issues surrounding different surface tension implementations.

\vspace*{0.2cm}


The two most popular approaches that deal with interfacial transport are the Volume-of-Fluid (VOF) method first developed by Hirt and Nichols \cite{hirt1981volume}, and the level set class of methods pioneered by Osher and Sethian \cite{osher1988fronts}. 
Each class of methods has its own set of merits (and demerits) relative to each other. Generally speaking, Volume-of-Fluid based methods display superior mass conservation whereas in terms of interface curvature computation, level set based methods hold an advantage. 
For a more detailed and nuanced evaluation of the comparitive advantages of interfacial transport methods, we refer the reader to the recent review by Mirjalili et al. \cite{mirjalili2017interface} on the given subject.     
An additional feature one has to consider in the context of Volume-of-Fluid based methods is that of conservative mass transport in 3D. 
Fluxes can be computed via algebraic transport schemes (generally less accurate), or by using geometric reconstructions in either Eulerian, Lagrangian or hybrid frameworks. 
The temporal integration of the fluxes could be carried out either as a series of one dimensional propagations along each of the spatial directions, termed as direction-split, or carried out in one single sweep, termed as multidimensional or unsplit. 
Direction-split methods are more intuitive and easier to develop (extension to 3D in particular), 
but generally suffer from lack of conservation (to the order of machine precision) when it comes to 3D.
A notable exception to the above point is the conservative algorithm originally developed by Weymouth \& Yue \cite{wy},
which maintains conservative direction-split VOF transport in 3D (subject to local CFL restrictions). 
Multidimensional (unsplit) methods have an advantage in that respect due to the fact that they are conservative by nature of their design, but are inherently more complicated to develop and implement, with no straightforward extension from 2D to 3D.    

\vspace*{0.2cm}

The first study to address the issue of consistency between mass and momentum transport was the seminal work of Rudman \cite{rudman1998volume}.
The fundamental hurdle in the implementation of mass-momentum consistent transport for staggered configurations of primary variables (pressure and velocity) is the inherent difficulty in reconstructing mass (defined on centered control volumes) and its corresponding fluxes onto the staggered control volumes on which momentum is defined. 
Rudman introduced the strategy of carrying out mass advection (although using algebraic flux reconstructions) on a grid twice as fine as that of momentum, thereby enabling a `natural' and intutive way to reconstruct mass and its fluxes onto staggered momentum control volumes. 
However, the method uses a VOF based convolution technique for curvature computation, which is neither consistent nor well-balanced.    
Bussmann et al. \cite{bussmann2002modeling} were able to circumvent the issue surrounding staggered grids altogether by using a collocated arrangement in the context of hexahedral unstructured meshes, coupled with an unsplit Eulerian flux computation method.    
The study though makes no mention of any surface tension model. 

\vspace*{0.2cm}

Level set based methods in the context of mass-momentum consistent transport were implemented first by Raessi and Pitsch \cite{raessi2012consistent}, followed by Ghods and Hermann \cite{ghods2013consistent}. 
In the former, the consistency problem is tackled by means of a semi-Lagrangian approach, computing geometric level set derived fluxes at two different time intervals, whereas in the latter, a collocated arrangement is used. 
Nonetheless, both methods face certain drawbacks, notably the applicability only to 2D in case of Raessi and Pitsch, as well as a lack of well-balanced surface tension models for both these methods.   
Recent advances concerning Volume-of-Fluid based methods that employ unsplit (conservative) geometric flux reconstructions were made by LeChenadec and Pitsch \cite{le2013monotonicity}, and later by Owkes and Desjardins \cite{owkes2017mass}. 
LeChenadec and Pitsch utilize a Lagrangian remap method in order to construct consistent mass-momentum fluxes for the staggered control volumes, while Owkes and Desjardins use mass advection on a doubly refined grid (same principle as Rudman) to achieve consistency.     
Although \cite{le2013monotonicity} implements a well-balanced surface tension model, the VOF convolution based curvature computation is not consistent. 
In case of \cite{owkes2017mass}, they use mesh-decoupled height functions to compute curvature while coupling it with a well-balanced surface tension model. 
However, their semi-Lagrangian flux computation procedure involving streak tubes and flux polyhedra are extremely convoluted in 3D.     

\vspace*{0.2cm}

Certain methods attempt to combine the qualities of both Volume-of-Fluid and level set methodologies (CLSVOF), as proposed in the works of Vaudor et al. \cite{vaudor2017consistent}, and more recently by Zuzio et al. \cite{zuzio2020new}. 
They both tackle the consistency issue by means of projecting the direction-split geometric fluxes onto a twice finer grid, which are subsequently recombined to reconstruct consistent fluxes for mass and momentum for the staggered control volumes. 
This approach allows them to bypass the requirement of conducting mass advection on a twice finer grid (as in the original Rudman method), thereby deriving the benefits of a sub grid without the added computational cost of doubly refined mass transport. 
In addition, both methods adopt well-balanced surface tension models with consistent level set based curvature estimation. 
However, the purported advantages of both these methods with regards to reduced computational costs is not quite evident, as additional complexities are introduced due to the projection (reconstruction) of fluxes onto a the twice finer mesh, which would not be necessary in the first place if mass transport had been carried out on the twice finer mesh itself. 
Patel and Natarajan \cite{patel2017novel} developed a hybrid staggered-collocated approach to solve the consistency issue on polygonal unstructured meshes, complemented with a well-balanced surface tension model. Nevertheless, the VOF advection is based on algebraic transport, not to mention the use of a VOF convolution based curvature computation, which is inherently not consistent. 
More recently, Nangia et al. \cite{nangia2019robust} developed a CSLVOF method for dynamically refined staggered Cartesian grids. They utilize Cubic Upwind Interpolation (CUI) schemes to reconstruct consistent mass and momentum fluxes on the staggered control volumes, using the information from the additional mass advection equation they solve alongside the level set function.   
However, the reconstruction of mass fluxes using CUI schemes are inherently algebraic, with their comparative advantage against fluxes computed via geometric constructions being an open question (refer to Mirjalili et al. \cite{mirjalili2017interface}) .
In our previous study \cite{caf2020}, the issue of consistent mass-momentum advection 
on staggered grids was tackles by using \textit{shifted fractions}.
Although in this approach the \textit{shifted} (mass and momentum) fields are transported in a 
conservative direction-split fashion, the method is not exactly conservative when it comes to momentum.  
A comparative summary of all the methods in existing literature can be found in Appendix A \ref{sec:append_a} .

\vspace*{0.2cm}

In the present study, we adopt the strategy of mass advection on a twice finer 
grid in order to enable consistent reconstruction of mass and momentum on 
the staggered control volumes, in the spirit of Rudman \cite{rudman1998volume}. 
A geometric representation of the interface within the Volume-of-Fluid framework is utilized
, along with geometric reconstruction of fluxes whose temporal integration is a carried out in a direction-split manner. 
The conservative direction-split mass (volume) transport algorithm of 
Weymouth and Yue \cite{wy} is extended to the direction-split transport of momentum. 
The implementations of the algorithms are developed on the free and open-source 
numerical platform called ``PARIS Simulator'' \cite{paris}, with the detailed descriptions 
of the general capabilities of the solver to be found in the reference provided.
The organization of the paper is quite simple, in section \ref{sec:goveqn} we describe our set of 
governing equations, followed by their numerical implementation in section \ref{sec:method}. 
In the numerical benchmarks section \ref{sec:bench} we assess the accuracy of our consistent and 
conservative method by means of canonical test cases which are well-established in existing literature. 
Finally we present the numerically challenging case of a raindrop falling in air. 
In the process, we evaluate the robustness, stability and accuracy of the present method compared to
the standard version of our method which does not ensure consistent mass-momentum transport. 
We wrap up this study with some concluding remarks and future perspectives.

\label{sec:intro}

\section{Governing Equations}
\label{sec:goveqn}
In this section, we describe our methodology behind modeling 
the dynamics of immiscible incompressible liquid-gas 
interfacial flows under isothermal conditions. 

\subsection{Navier Stokes with Surface Tension}

We use the one-fluid formulation for our system of governing equations, thus solving
the incompressible Navier-Stokes equations throughout the whole domain including regions
of variable density and viscosity, which itself depend on the explicit
location of the interface separating the two fluids.
In the absence of mass transfer, the velocity field is continuous across
the interface at the incompressible limit, with the interface evolving 
according to the local velocity vector. Thus, the equations are 


\begin{align}
        \frac{\partial \rho}{\partial t} + \nabla\cdot \left(\rho\boldsymbol{u}\right) &= 0 \, , \label{mass} \\
	\frac{\partial}{\partial t} \left(\rho\boldsymbol{u}\right) + \nabla \cdot \left(\rho\boldsymbol{u}\otimes\boldsymbol{u}\right)  &= -\nabla p + \nabla \cdot \left(2 \mu \boldsymbol{D}\right) + \sigma \kappa \delta_{s}\boldsymbol{n} + \rho \boldsymbol{g} \, , 
\label{nseqn}
\end{align}

with $\rho$ and $\mu$ being the density and dynamical viscosity respectively.
The volumetric sources are modeled by the acceleration $g$, and the
deformation rate tensor $\boldsymbol{D}$ used to model the viscous stresses defined as

\begin{align}
        \boldsymbol{D} = \frac{1}{2}\left[\nabla \boldsymbol{u} + \left(\nabla \boldsymbol{u}\right)^{T}\right] . 
\end{align}

The term $\sigma \kappa \delta_{s}\boldsymbol{n}$ models the surface tension forces in the
framework of the continuum surface-force (CSF) method \cite{csfmodel}. The normal vector to the interface
is $\boldsymbol{n}$, $\sigma$  the coefficient of surface tension and $\kappa$ the
local interfacial curvature. The operator $\delta_{s}$ is the Dirac delta function,
the numerical approximation of which allows us to map the singular surface force distribution
along the interface onto its volumetric equivalents.
At the incompressible limit, the advection of mass given by equation \ref{mass} can be
treated as equivalent to that of the advection of volume.

\subsection{Material Properties} 

Within the framework of interface capturing schemes, the
temporal evolution of the interface separating the two fluids
can be tracked by the advection equation 

\begin{align}
        \frac{\partial \chi}{\partial t} + \boldsymbol{u}\cdot \nabla\chi = 0 \, , 
\label{chi}
\end{align}

where $\chi$ is the phase-characteristic function, that has different values
in each phase. Generally, $\chi$ is assigned a value of $0$ in one phase and $1$ in the other. 
Mathematically, the function $\chi$ is equivalent to a Heaviside function in space and time.
At the macroscopic length scales under consideration, the interface evolution
as described by equation \ref{chi} is modeled as having infinitesimal thickness
under the continuum hypothesis. The coupling of the interfacial evolution with
the equations of fluid motion as described in \ref{mass} and \ref{nseqn} is provided by 

\begin{align}
        \rho &= \rho_{1}\chi + \left(1 - \chi\right)\rho_{2} \label {rho_chi} \, , \\
        \mu  &= \mu_{1}\chi  + \left(1 - \chi\right)\mu_{2} \, , 
  \label{mu_chi}
\end{align}

where $\rho_{1}$, $\rho_{2}$ are the densities of fluids 1 and 2 respectively,
likewise for viscosities $\mu_{1}$ and $\mu_{2}$. For certain flow configurations,
it might be beneficial to opt for a weighted harmonic mean description of the
variable dynamic viscosity \cite{harmonic_mean}, instead of the weighted arithmetic mean as in equation \ref{mu_chi}.


\section{Numerical Methodology}
\label{sec:method}
The governing equations are numerically solved using finite volume discretizations on uniform Cartesian grids, 
utilizing state of the art methods in interfacial reconstruction coupled with geometric
transport of the corresponding fluxes, curvature computation and surface tension modeling. 
A detailed exposition of our class of mass-momentum consistent numerical methods is provided,
which are specifically designed to circumvent or suppress the uncontrolled and rapid
growth of numerical instabilities that arise when dealing with flows entailing marked density contrasts .

\subsection{Volume-of-Fluid Method}
\label{subsec:vof}
Our numerical studies are based on the Volume-of-Fluid methodology. 
We refer to the discontinuous approximation to the Heaviside function
$\chi$ as the volume fraction field or colour function interchageably, which 
is defined below in the context of finite volume discretization as  

\begin{align} 
        C_{i,j,k}\left(t\right) = \frac{1}{\Delta V} \displaystyle\int_{\Delta V} \chi(\boldsymbol{x},t) \,d\boldsymbol x \, , 
	\label{vof_basic}
\end{align}

\begin{figure}[h!]
\includegraphics[width = 1.0\textwidth]{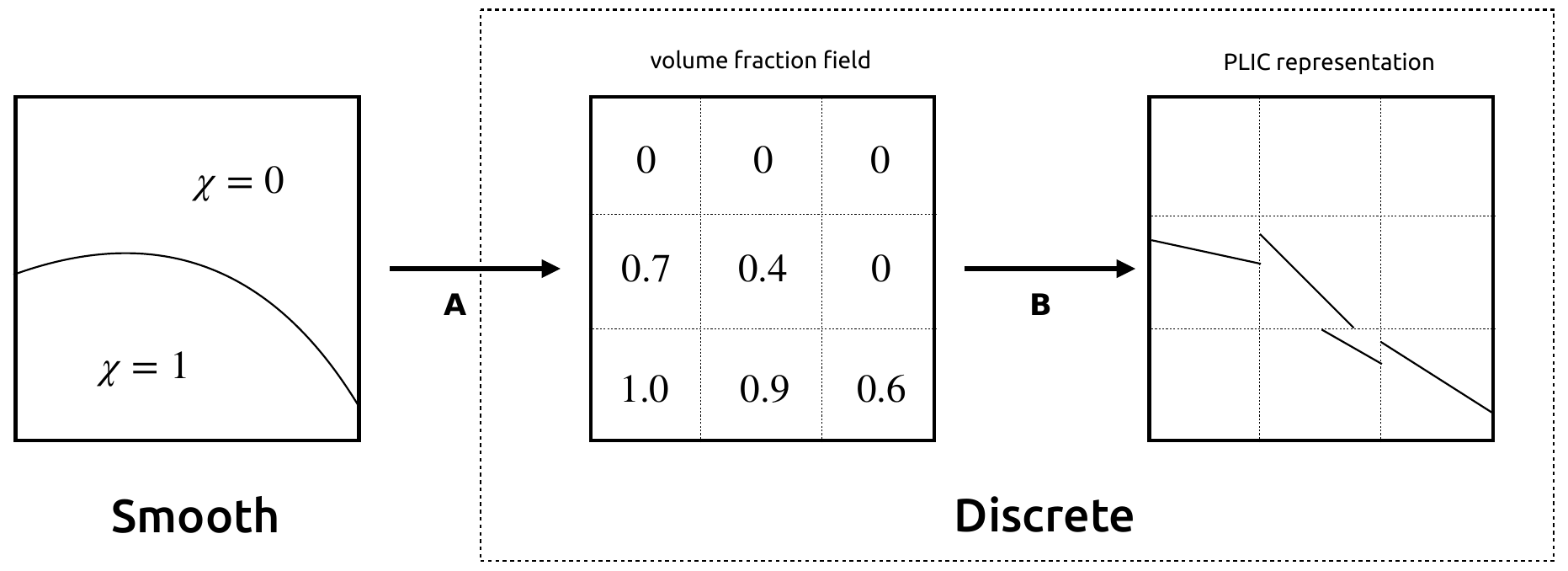}
\centering
\caption{ Exlicit definition of the interface location using the volume-of-fluid approach. 
The smooth interfacial representation is discretely represented on the Cartesian grid 
using the volume fraction (VOF) field, which is defined in \eqref{vof_basic}.
The interface is geometrically reconstructed using the volume fraction field 
as piecewice continuous line segments (PLIC), details of which can be found in \cite{zaleskibook, gueyffier}. 
}
\label{vof_discrete}
\end{figure}

where $C$ is the colour function with its values lying between $0$ and $1$, 
$i$,$j$ and $k$ being the indices to the corresponding discretized control volume $\Delta V$.  
There are two steps involved in the VOF method, 
the reconstruction of the interface, and its subsequent propagation (advection). 
The interface is represented by disjointed line segments under 
the PLIC (piecewise linear interface construction) framework. 
Such reconstructions involve the determination of interface normals 
using the Mixed Youngs Centered method (refer to \cite{zaleskibook,gueyffier}). 
Once the geometric PLIC reconstructions have been carried out,
the interface segments are advected using the the velocity field.
This entails computation of the fluxes in a geometric manner, 
and subsequent integration of the fluxes in a direction-split manner. 
Towards that objective, we employ the Eulerian implicit scheme originally
developed by Weymouth and Yue \cite{wy} in order to specifically tackle 
the problem of discrete conservation when it comes to direction-split schemes.
The scheme is basically a forward Eulerian method with respect to the temporal
integration of fluxes i.e. the fluxes are computed as the quantity of fluid 
entering or exiting a given control volume through its fixed surfaces as shown in figure 
We start with the advection of the interface position by the velocity field $\boldsymbol{u}$, 
using the conservative form of \eqref{chi} as -  

\newcommand{\dert}[1]{\frac{\partial #1}{\partial t}}
\newcommand{\cijk}{C_{i,j,k}}
\newcommand{\pijk}{\phi_{i,j,k}}
\newcommand{\X}{\boldsymbol{x}}
\newcommand{\N}{\boldsymbol{n}}
\newcommand{\rijk}{\rho_{i,j,k}}
\newcommand\mijk{(\rho u_{q})_{i,j,k}}

\begin{align}
        \dert \chi + \nabla \cdot (\boldsymbol{u} \chi) = \chi \nabla \cdot \boldsymbol{u} \,. 
        \label{heaviside}
\end{align}

As one can observe, the ``compression'' term on the right hand side of 
\eqref{heaviside} equals to zero in the context of incompressible flows without mass
transfer, but it is important to keep this term in our numerical formulation within
the direction-split framework. Integrating this equation in time after carrying out 
spatial discretization, one obtains 

\begin{align}
{\cijk^{n+1} - \cijk^{n}} = - \sum_{\textrm{faces} \, f} F^{(C)}_f + \int_{t_n}^{t_{n+1}}
        {\textrm{d}}t \int_\Omega  \chi \nabla \cdot \boldsymbol{u} \,  {\textrm {d}}\X   \,,
\label{sumf}
\end{align}

where the first term on the right-hand side is the summation over the cell faces $f$
of the fluxes $F^{(c)}_f$. These fluxes of $(\boldsymbol{u} \chi)$ are computed 
via geometrical reconstructions, and not via high-order non-linear interpolation 
schemes which are the standard in the absence of discontinuities.
The definition of the geometric fluxes in mathematical form is 

\begin{align}
        F_f^{(C)} = \int_{t_n}^{t_{n+1}} {\textrm {d}}t \int_{f} u_f(\X,t) \chi(\X,t) \, {\textrm{d}}\X \,,
\label{faceint}
\end{align}

where $u_f = \boldsymbol{u}\cdot \N_f$ is the component of the 
velocity normal to the control surface $f$.
Directional splitting results in the decomposition of equation (\ref{sumf})
into three equations, one for each advection substep as 

\begin{align}
{\cijk^{n,l+1} - \cijk^{n,l}} = - F^{(C)}_{m-} - F^{(C)}_{m+}
+ c_m \partial_{m}^h u_m .  
\label{sumf2}
\end{align}

\begin{figure}[!h]
\includegraphics[width=0.7\textwidth]{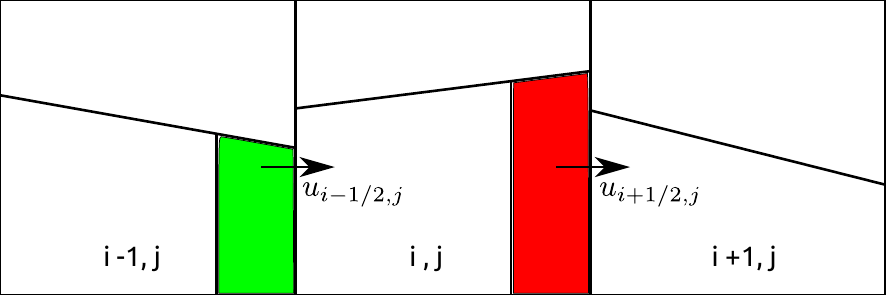}
\centering
\caption{ A 2D schemetic of the Eulerian (geometric) flux calculation  
using the Weymouth-Yue \cite{wy} algorithm for the advection substep
along the horizontal direction, with the interface reconstructed 
using the volume fraction field at the start of the substep.
The colour fraction of the central cell ($i,j$) is updated during this substep
through the addition of the fluxes (coloured regions), with the green polygon corresponding
to the volume entering the cell $i,j$ from the $i-1,j$ and the red one corresponding to
that exiting $i,j$ into $i+1,j$. The geometric flux calculations are made on the basis
face centered velocities of the cell $i,j$ at the start of the advection substep.}
\label{fig:wy}
\end{figure}

The superscript $l=0,1,2$ is the substep index, i.e.
$\cijk^{n,0} = \cijk^{n}$ and $\cijk^{n,3} = \cijk^{n+1}$.
The face with subscript $m-$ is the ``left'' face in direction $m$ with
$F^{(C)}_{m-} \ge 0$ if the flow is locally from right to left. 
A similar reasoning applies to the ``right'' face $m+$. 
After each advection substep (\ref{sumf2}), the interface is reconstructed
with the updated volumes $\cijk^{n,l+1}$, then the
fluxes $F^{(C)}_{f}$ are computed for the next substep.
Importantly, we have approximated the compression term in (\ref{sumf}) by 

\begin{align}
        \int_{t_n}^{t_{n+1}}  {\textrm {d}}t \int_\Omega  \chi \partial_m u_m  {\textrm {d}}\X \simeq  c_m
 \partial_{m}^h u_m .  
 \label{comp}
\end{align}

One must note that in the above equation there is no implicit summation carried out
over $m$, and the superscript $h$ denotes the spatial discretization of the operator. 
On the right hand sides of (\ref{sumf2}) and (\ref{comp})
the flux terms $F_{f}^{(c)}$ and the
 partial derivative $\partial_{m} u_m$ must
be evaluated using identical discretized velocities.
The expression $\partial_{m}^h u_m$ is a finite volume approximation of the
spatial derivative corresponding to the $m$th component
of the velocity vector along direction $m$, and
the ``compression coefficient'' $c_m$ approximates the color fraction.
The exact expression of this coefficient depends on the advection method
and it also entails the desirable property of $C$-bracketing
(preservation of $0 \le \cijk \le 1$).
The possible dependency of $c_m$ on the Cartesian direction
corresponding to the advection substep might render the discrete sum 
$\sum_m c_m \partial_{m}^h u_m$ to be non-zero, even if the flow is solenoidal.  
The primary appeal of the Weymouth-Yue method lies in the subtle but important tactic 
used to estimate the prefactor to the compression term, with its definition being  

\begin{align}
c_{m} = H \left( C_{i,j,k}^{n,0} - 1/2 \right) \, , 
\label{wy_cond}
\end{align}

where $H$ is a one-dimensional Heaviside function. This renders the compression coefficient
independent of the direction of the advection substep, consequently enabling the
three discrete directional divergences to sum up to zero i.e. $\sum_m c_m \partial_{m}^h u_m = 0$.
For a formal proof, we refer the reader to the appendix of the original study \cite{wy}, 
which demonstrates that volume conservation is guaranteed, given certain local CFL restrictions. 
From a practical point of view,  we can only ensure that they sum up to the 
tolerance of the Poisson solver (convergence criteria usually set between 
$10^{-3} - 10^{-6}$), and furthermore, the sum is ultimately 
limited by the level of machine precision ($\sim 10^{-14} - 10^{-17}$)
available for representing floating point numbers. 
In the subsections that follow, we detail the extension of this exactly 
conservative direction-split transport framework to the advection of mass and momentum. 


\subsection{Consistent Mass-Momentum Transport}
In order to achieve consistent transport of the discontinuous fields 
of mass and momentum, we start by trying to understand the advection
of a generic \textit{conserved} scalar quantity $\phi$ by a continuous velocity field

\begin{align}
        \dert \phi + \nabla \cdot \left( \phi \, \boldsymbol{u} \right)  = 0 \, , 
\label{phiconv}
\end{align}

where the field $\phi$ is smoothly varying except
at the interface position, where it may be discontinuous.
The smoothness of the advected quantity away from the interface
is verified for fields such as density ($\rho$) and  momentum ($\rho \boldsymbol{u}$). 
A major theme of this study lies in the search for a scheme that
propagates this discontinuity ($\phi$) at the same speed as
that corresponding to the advection of volume fraction ($C$).
Temporal integration of the spatially discretized version of \eqref{phiconv} gives us 

\begin{align}
\pijk^{n+1} - \pijk^{n} = - \sum_{\textrm{faces}\, f} F^{(\phi)}_f \, . 
\label{sumfp}
\end{align}

The sum on the right-hand side is the sum over faces $f$ of cell $i,j,k$
of the fluxes $F^{(\phi)}_f$ of $\phi$. The flux definitions are given by  

\begin{align}
        F_f^{(\phi)} = \int_{t_n}^{t_{n+1}} {\textrm {d}}t \int_{f} u_f(\X,t) \,\phi(\X,t) \,  
        {\textrm {d}}\X \, .
\label{pfaceint}
\end{align}

In order to ``extract'' the discontinuity we introduce the interface Heaviside function $ \chi(\X,t) $, 
thus giving us 

\begin{align}
F_f^{(\phi)} =
\int_{t_n}^{t_{n+1}} {\textrm d}t \int_{f} \left[ u_f  \,\chi \,\phi  +  u_f \,(1-\chi) \,\phi \right]
\, {\textrm d}\X .
\label{fluxphi}
\end{align}

Therefore, the flux can be decomposed into two components as 

\begin{align}
F_f^{(\phi)} =
\bar \phi_1 \int_{t_n}^{t_{n+1}} {\textrm d}t \int_{f} u_f \,\chi \,{\textrm d}\X +
\bar \phi_2 \int_{t_n}^{t_{n+1}} {\textrm d}t \int_{f} u_f \,(1-\chi) \,{\textrm d}\X \,,
\label{barphi}
\end{align}

where the face averages $\bar \phi_s$, $s=1,2$, are defined as 

\begin{align}
\bar \phi_s = \frac{\int_{t_n}^{t_{n+1}} {\textrm d}t \int_{f} \phi \,u_f \,\chi_s
{\textrm d}\X}{\int_{t_n}^{t_{n+1}} {\textrm d}t \int_{f}
u_f \,\chi_s\,{\textrm d}\X} \,,
\label{barphi2}
\end{align}

and $\chi_1=\chi$, $\chi_2= 1-\chi$.
The total flux can be rearranged as a sum of the constituents
corresponding to the different ``fluids''

\begin{align}
F_f^{(\phi)} = \bar \phi_1 \,F_f^{(C)} +  \bar \phi_2 \,F_f^{(1-C)} \, .
\label{fluxphi12}
\end{align}

\subsubsection{Mass Transport}
Moving forward, the density field $\rho(\X,t)$ follows the
temporal evolution of the generic conserved quantity
(\ref{phiconv}) by simply setting $\phi = \rho$.
At the incompressible limit the velocity field is
solenoidal (divergence-free), with constant densities
in each phase. We can extract the density trivially from the integrals
\eqref{barphi2} to \textit{exactly} obtain $\bar \rho_s = \rho_s$.
The flux definitions corresponding to $\rho$ thus become

\begin{align}
F_f^{(\rho)} = \rho_1 F_f^{(C)} +  \rho_2 F_f^{(1-C)} \, . 
\label{fluxrho}
\end{align}

Using the above definitions for density fluxes,
we use our VOF method to construct fluxes of the color
function in order to obtain conservative transport for $\rho$. 
In principle, the discretization of \eqref{phiconv} while setting $\phi = \rho$
should result in conservation of total mass during the advection step.
However, as we can observe, the discrete transport of the density field \eqref{sumfp} 
does not have the ``compression term'' which is present in the discrete transport of volume fraction \eqref{sumf2}. 
Thus we have to make several adjustments in order to render the transport of mass consistent
with that of volume fraction at the discrete level. 
We start by adding the divergence term on the right hand side of the 
continuous form of mass transport \eqref{phiconv}, which gives us 

\begin{align}
\dert \phi + \nabla \cdot ( \phi \,\boldsymbol{u})  = \phi \, \left(\nabla \cdot \boldsymbol{u}\right) \, . 
\label{phiconv2}
\end{align}

This equation can be decomposed into advection substeps corresponding to
the direction-split integration framework as 

\begin{align}
{\pijk^{n,l+1} - \pijk^{n,l}} = - F^{(\phi)}_{m-} - F^{(\phi)}_{m+}
+ \Big( \tilde \phi_1^m c^{(1)}_m + \tilde \phi_2^m c^{(2)}_m \Big) \, \partial_{m}^h u_m \, .
\label{sumfpconsistent}
\end{align}

The term $c^{(1)}_m=c_m$ is the compression coefficient corresponding
to the VOF advection method, while $c^{(2)}_m = 1 -c_m$ corresponds to that 
of the symmetric color fraction  $1 - C$.
The fluxes $F^{(\phi)}_{m\pm}$ follow the definition given in \eqref{fluxphi12}
, while the cell averages $\tilde \phi_s^m$ are defined as

\begin{align}
\tilde \phi_s^m = \frac{\int_{t_n}^{t_{n+1}} {\textrm{d}}t \int_{\Omega}  \phi  \chi_s
\partial_{m}^h u_m  \,  {\textrm{d}}\X}
{\int_{t_n}^{t_{n+1}} {\textrm{d}}t \int_{\Omega} \chi_s  \partial_{m}^h u_m \,{\textrm{d}}\X} \, . 
\label{cellphi2}
\end{align}

The direction-split integration operation for $\rho$ is rewritten as  

\begin{align}
{\rijk^{n,l+1} - \rijk^{n,l}} = - F^{(\rho)}_{m-} - F^{(\rho)}_{m+} + C_m^{(\rho)}\,,
\label{sumfrho}
\end{align}

where the fluxes are given by (\ref{fluxrho}), and the corresponding compression term is

\begin{align}
C_m^{(\rho)} =  \left( \rho_1 c^{(1)}_m + \rho_2 c^{(2)}_m \right) \,\partial_{m}^h u_m \, . 
\label{central}
\end{align}

The above equations (\ref{sumfrho},\ref{central}) ensures that the discrete directional divergence terms of the mass \eqref{sumfp}
and volume fraction \eqref{sumf2} transport equations are consistent at the discrete level.  
As we have discussed before, the Weymouth-Yue method ensures that the 
``directional'' compression terms eventually cancel upon summation over the substeps, 
therefore resulting in the conservation of mass at the same accuracy as 
the discrete incompressibility condition $\sum_{m=1}^3 \partial_{m}^h u_m=0$ is verified.  

\subsubsection{Momentum Transport}
Thus far, we have made the direction-split (conservative) advection of volume consistent with that of density (mass). 
We now consider momentum advection within the framework of the conserved scalar transport, 
by setting $\phi=\rho u_q$, where $q=1,2,3$ is the component index.
Using the definition given in \eqref{barphi2}, we obtain the expression 

\begin{align}
\overline{\rho u_q}_s = \rho_s \bar u_{q,s} \,,
\end{align}

where $\bar u_{q,s}$ is termed as the ``advected interpolated velocity'',
whose precise definition is given as 

\begin{align}
        \bar u_{q,s} =  \frac{\int_{t_n}^{t_{n+1}} {\textrm{d}}t \int_{f}  u_q u_f  \chi_s  \, 
	{\textrm{d}}\X}{\int_{t_n}^{t_{n+1}} {\textrm{d}}t \int_{f}  u_f  \chi_s\,{\textrm{d}}\X} \, .
	\label{barudef}
\end{align}

As a reminder, the subscript $s$ denotes the phase or fluid,
and $f$ represents the normal components defined on the face centers.
Thus, we obtain the expression for the direction-split integration of the momentum as 

\begin{align}
{\mijk^{n,l+1} - \mijk^{n,l}}  = - F^{(\rho u)}_{m-} - F^{(\rho u)}_{m+}
 + \Big( \rho_1 \tilde u_{q,1}^m  c^{(1)}_m +  \rho_2 \tilde u_{q,2}^m c^{(2)}_m
 \Big) \, \partial_{m}^h u_m \,,
\label{sumfrou}
\end{align}

where the momentum fluxes are constructed using 

\begin{align}
 F^{(\rho u)}_{f} =  \rho_1 \,\bar u_{q,1}  \,F^{(C)}_{f}  +
 \rho_2 \,\bar u_{q,2}  \,F^{(1-C)}_{f} \,.
\end{align}

The expression for the ``central interpolated velocity'' corresponding to the
averages $\tilde \phi_s^m$ of \eqref{cellphi2} is 

\begin{align}
        \tilde u_{q,s}^{m} = \frac{\int_{t_n}^{t_{n+1}} {\textrm{d}}t \int_{\Omega} u_q \, H_s \,
        \partial_{m}^h u_m   \,  {\textrm{d}}\X}
        {\int_{t_n}^{t_{n+1}} {\textrm{d}}t \int_{\Omega} H_s \, \partial_{m}^h u_m \,{\textrm{d}}\X} \,.
\label{tildeudef}
\end{align}

The superscript $m$ is intentionally omitted  
for the velocities $\tilde u_q^m$ in order to avoid
cumbersome and complicated notations. 
As a reasonable approximation, we choose to put 
$\bar u_q =  \bar u_{q,1} = \bar u_{q,2}$ for the 
``advected interpolated velocity'' and 
$\tilde u_q =  \tilde u_{q,1} = \tilde u_{q,2}$
for the ``central interpolated  velocity''. 
To add clarity to the notion of ``central interpolated velocity'',
one can interpret this as the face-centered interpolations of the velocity field
(component normal to control surface), which is required to compute the volume fluxes. 
This leads us to an important simplification given by   

\begin{align}
 F^{(\rho u)}_{f} = \bar u_q F^{(\rho)}_{f} \label{frou} \, .
\end{align}

This model is central in our effort towards consistent mass-momentum transport. 
Therefore, the advection substep for the momentum can finally we written as 

\begin{align}
{\mijk^{n,l+1} - \mijk^{n,l}} =  -\bar u_q  F^{(\rho)}_{m-} - \bar u_q  F^{(\rho)}_{m+}
+ \tilde u_q C_m^{(\rho)} \,,
\label{sumfmom2}
\end{align}

where the density fluxes are defined in \eqref{fluxrho} and the compression term
$C^{(\rho)}$ in \eqref{central}.
In the above expression, the face-weighted
average velocities $\bar u_q$ are defined
using (\ref{barudef}) on the corresponding
left face $m-$ or right face $m+$.

As we have referred to in section \ref{subsec:vof}, the compression coefficient is independent
of the directional substep due to the adoption of the Weymouth-Yue algorithm.
The final expression for the compression coefficient becomes 

\begin{align}
	C_m^{(\rho)} =  \Big( \rho_1 C^{n,l} + \rho_2 (1-C^{n,l} ) \Big) \,\partial_{m}^h u_m  \,.
\label{central2}
\end{align}

Since there is no bracketing on any velocity component,
we take $\tilde u_q = u_q^{n}$, which is independent of the substep $l$.
The final expression after cancellation of the compression terms,
having undergone three advection substeps \eqref{sumfmom2} is

\begin{align}
        {\mijk^{n,3} - \mijk^{n}} =  - \sum_{\textrm{faces} \, \textrm{f}}  \bar u_{q}  F^{(\rho)}_{f} \,.
\label{sumfmomtotfracstep}
\end{align}

Therefore, the extension of the Weymouth-Yue algorithm for exact mass (volume)
conservation has been extended to incorporate the transport of momentum, 
thus ensuring consistency at the discrete level between mass and momentum transport. 
The manner in which the velocity interpolations concerning the weighted averages 
$\bar u_q$ and $\tilde u_q$ are carried out (near the interface and in the bulk), 
are identical to that covered in our previous study \cite{caf2020}, hence in the interest of brevity
the reader can refer to the appendix of the aforementioned reference. 

\subsection{Spatio-Temporal Discretization}

\begin{figure}[h!]
\begin{center}
\includegraphics[width=\textwidth]{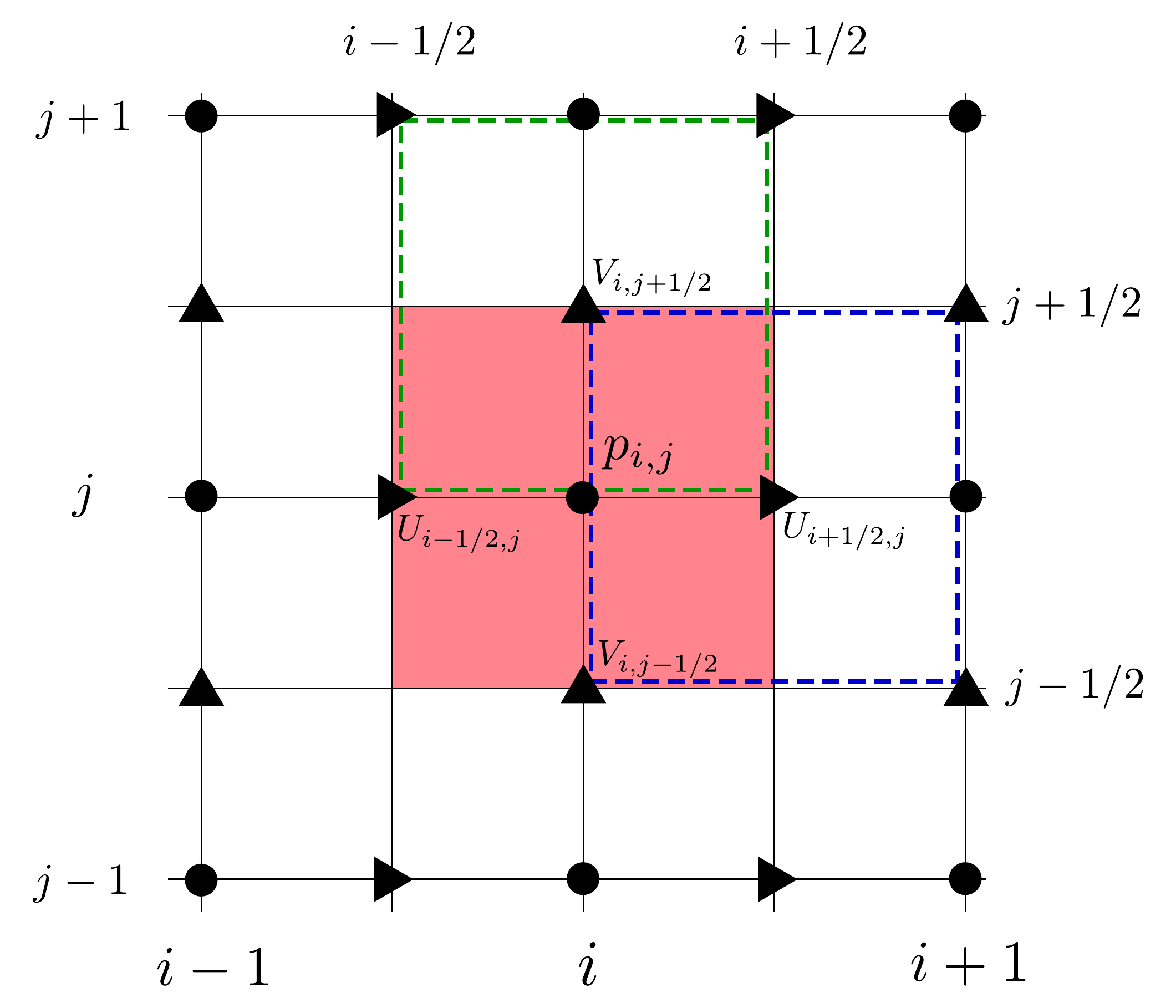}
\end{center}
\caption{A 2D schematic of the staggered spatial configuration of the
pressure and velocity variables.
The pressure $p_{i,j}$ is based on the center of its control volume (light red);
the horizontal velocity component $U_{i+1/2,j}$ is defined in the middle of the
right edge of the pressure control volume and centered on its control volume
(blue dashed box); the vertical velocity component $V_{i,j+1/2}$ is defined in the
middle of the top edge and centered on its corresponding control volume (green dashed box).
}
\label{stag-grid}
\end{figure}

We start by describing the spatial arrangement of our
primary variables i.e. pressure ($p$) and velocity $\boldsymbol{u}$,
where $U$,$V$ and $W$ represent the $x$,$y$ and $z$ componenets of $\boldsymbol{u}$ respectively.
The control volume is in the form of a cube (3D) or a square (2D).
The pressure and velocity variables are defined in a staggered arrangement,
which is illustrated in Figure \ref{stag-grid}.
The use of staggered control volumes has the advantage of
suppressing neutral modes (pressure oscillations) often observed in
collocated methods but leads to more complex discretizations
(refer to \cite{zaleskibook} for a detailed discussion) .
In order to describe the overall numerical algorithm for the one-fluid
Navier-Stokes equations with variable density and viscosity,
we choose to the reframe our equations in a more convenient operator form, given as  

\begin{align}
   \frac{\partial}{\partial t} \left( \rho \boldsymbol{u} \right) = L\left( \rho,\boldsymbol{u} \right) - \nabla p \,.
\end{align}

The operator $L$ in the above expression can be decomposed as 

\begin{align}
        L = L_{\textrm{adv}} + L_{\mu} + L_{\sigma} + L_{\textrm{g}} \,,
   \label{opr}
\end{align}

where the $L_{\textrm{adv}}$ represents the conservative advection,
$L_{\mu}$ represents the diffusive forces generated by viscous stresses,
$L_{\sigma}$ represents the capillary forces arising from the surface tension model
and finally $L_{g}$ represents the volumetric (body forces) source term.
We apply the spatially discretized versions of these operators (denoted by the superscript $h$)
onto the primary variables ($c,\boldsymbol{u}$), and march forward in time using a small,
possibly variable time-step $\tau$ such that $t_{n+1} = t_{n} + \tau$.
The volume fraction ($c$) is defined on the sub-grid (covered in the next subsection),
whereas pressure and momentum are defined at the course grid level.
In the first part of the algorithm, the volume fraction field $c^{n}$ is updated to the next timestep
, with the superscript $n$ signifying discretization in time.
The operation can be written as 

\begin{align}
        c^{n+1} = c^{n} + \tau L^{h}_{\textrm{vof}}\left( c^{n},\boldsymbol{u}^{n}\right) \,.
\label{cvof_update}
\end{align}

The subscripts $i,j,k$ are dropped in equations \eqref{cvof_update}
and \eqref{mom_update}, with the understanding that the operators in
equation \ref{opr} apply identically to all control volumes.
The temporal evolution of the volume fraction field represented above by the
operator $L^{h}_{\textrm{vof}}$ is in accordance with the Eulerian implicit Weymouth-Yue
advection scheme, as described in section \ref{subsec:vof}.
Once we have obtained the updated field $c^{n+1}$, we can move on to
the temporal update of our momentum field given by

\begin{align}
        \rho^{n+1}\cdot \boldsymbol{u}^{*} &= \rho^{n}\cdot \boldsymbol{u}^{n} + \tau L^{h}_{\textrm{adv}}\left( C^{n},\boldsymbol{u}^{n} \right) + \nonumber \\  
                                      & \tau \left[ L^{h}_{\mu}\left(C^{n+1},\boldsymbol{u}^{n}\right) + L^{h}_{\sigma}\left(C^{n+1}\right) + L^{h}_{\textrm{g}}\left(C^{n+1}\right)\right] \,.
\label{mom_update}
\end{align}

The $C$ field is defined on the coarse grid, reconstructed 
using information from the sub-grid volume fraction ($c$) field.
In regions of constant density. the advection operator $L^{h}_{\textrm{adv}}$
is implemented using higher order spatial schemes
coupled with a choice of non-linear flux limiters such as
QUICK, ENO, WENO, Superbee, Verstappen and BCG.
These high-order spatial schemes are based on well established methods
developed to deal with hyperbolic conservation laws, for more details refer to the
studies of Leveque \cite{flim_1} and Sweby \cite{flim_2}.
For control volumes in the vicinity of the interface location, we revert to lower order
variants of these schemes due to the sharp jumps in the 
material properties across the interface (refer to \cite{caf2020}).
The functionality of the operator $L^{h}_{\textrm{adv}}$ near the interface is tighly coupled
to that of $L^{h}_{\textrm{vof}}$ from equation \ref{cvof_update},
thus representing the consistent volume-mass-momentum transport scheme which
was covered in the preceding section.

\subsection{Sub-grid Method}

The development of the consistent transport schemes should be integrated into
the broader context of our numerical algorithm which deals with the
coupling between the conservative formulation of the
one-fluid Navier-Stokes equations and the geometric transport of the interface.
In our particular approach, the difficulty associated with consistent 
transport on staggered control volumes is resolved by advecting  
the volume fraction (mass) on a twice finer grid,  
,very much in the spirit of Rudman's \cite{rudman1998volume} original work. 
In the method developed in our previous study \cite{caf2020} which we refer to as the 
\textit{shifted fractions} method, flux computations on the staggered 
cells necessitate another round of interfacial reconstructions based on the shifted 
volume fraction field after each advection substep, whereas the sub-grid volume fraction information enables us  
to circumvent the reconstructions and obtain the fluxes directly using simple surface integrals.
The sub-grid method allows us to evade complications that arise due to
the parallel evolution of two different volume fraction fields by conducting 
volume fraction advection solely on the twice refined grid.
From this point onwards, we use the terms sub-grid and fine grid interchangeably. 
An additional advantage of the having the sub-grid volume fraction field is that 
it facilitates a more ``natural'' computation of not only the staggered 
volume fraction field, but more importantly its fluxes. 
In contrast to the \textit{shifted fractions} method in which one can use both Lagrangian and Eulerian
fluxes, the implementation of the sub-grid lends itself compatible only with an 
Eulerian flux computation method (e.g. Weymouth-Yue scheme). 
Considering the overall approach, the \textit{key differences} between our 
present algorithm and the implementations in the 
original works of Rudman and Weymouth \& Yue are : 

\begin{description}
        \item[Rudman (1998)\cite{rudman1998volume} ] The use of geometrical 
                mass (volume fraction) transport instead of the algebraic transport in 
                the original method. 
        \item[Weymouth \& Yue (2010) \cite{wy}] The extension of the 
                original method for conservative direction-split mass transport to the momentum field, 
                culminating in the discrete consistency between mass 
                momentum transport. 
\end{description}

\subsubsection{Spatial Configuration : Coarse \& Sub-Grid Variables}

\begin{figure}[h!]
\includegraphics[width = 1.0\textwidth]{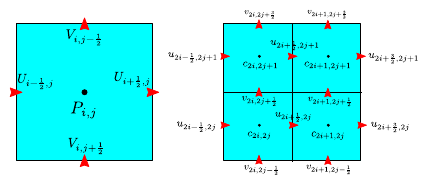}
\centering
\caption{A 2D schematic of the arrangement of primary variables
on the coarse and sub-grid levels.
The control volume on the left hand side is identical to the red colored 
square in Fig. \ref{stag_grid}, corresponding to the grid level for the momentum and pressure. 
Thus, the left hand side control volume corresponds to the coarse grid, while the
one on the right corresponds to the sub-grid .
Superposition of these two figures, one on top of each other,
represent the spatial relationships between the two sets of variables.
The velocities on the sub-grid level are simple first
or zeroth order interpolations (\eqref{sub_ops}) of the coarse grid velocities.
}
\label{sub_grid}
\end{figure}

To start, we describe the spatial orientation of the different
variables on the two refinement levels of the grid.
At the coarse grid level, the pressure and velocity fields are
defined in a staggered configuration, with pressures at the cell
centers (centroids in 3D) and velocities on the cell face centers.
In Fig. , we demonstrate the arrangement of variables,
which one can easily extrapolate to 3D during implementation,
but for the sake of clarity we illustrate its 2D equivalent.
The volume fraction $C$ is only defined on a grid that is twice as fine than
that of the pressure/velocity grid, which we refer to as the sub-grid.
Therefore in 3D, each cubic control volume is divided into eight constituent
smaller cubic volumes at the centroids of which, the volume fraction field is centered.

\subsubsection{Multiscale Coupling : Restriction \& Prolongation Operations}

As is paradigmatic for dual-grid methods in the context of
Navier-Stokes solvers, we carry out mass transport
at the sub-grid level, and momentum transport at the coarse level.
In pressure-projection based methods such as ours, the predominant
bottleneck in terms of computational speed is the
iterative solution to the discrete Poisson equation.
The problem size of such elliptical
partial differential equations is governed by
the number of cells/points discretizing the velocity (or pressure) field.
If we choose to solve the problem on a mesh with
twice the resolution in 3D, that would lead to an 8 fold increase
in the problem size for the discrete pressure-Poisson problem.
Therefore, solving the mass advection equation on the fine grid
allows us not only to obtain more accurate solutions to the flow physics
involved, but also enables us to avoid the significantly higher
computational costs associated with solving a Poisson problem
at a twice finer resolution.

\paragraph{\textbf{Prolongation}}

In order to advect the volume fraction at the sub-grid level,
we need to reconstruct a velocity field at that level of resolution using
information from the coarse grid velocity field.
The prolongation operator has to be used at the start of each new time step,
in order to compute the sub-grid velocity field.
The relationships between the coarse and fine grid discrete velocity fields as 
a result of the prolongation operator are given below  

\begin{align}
          u_{2i-\frac{1}{2},2j} = u_{2i-\frac{1}{2},2j+1} &= U_{i-\frac{1}{2},j} \nonumber \,, \\
            u_{2i+\frac{3}{2},2j} = u_{2i+\frac{3}{2},2j+1} &= U_{i+\frac{1}{2},j} \nonumber \,, \\
          u_{2i+\frac{1}{2},2j} = u_{2i+\frac{1}{2},2j+1} &= \left(U_{i+\frac{1}{2},j} + 
	  U_{i-\frac{1}{2},j}\right) / 2 \nonumber \,, \\
          v_{2i,2j-\frac{1}{2}} = v_{2i+1,2j-\frac{1}{2}} &= V_{i,j+\frac{1}{2}} \nonumber \,, \\
            v_{2i,2j+\frac{3}{2}} = v_{2i+1,2j+\frac{3}{2}} &= V_{i,j-\frac{1}{2}} \nonumber \,, \\
          v_{2i,2j+\frac{1}{2}} = v_{2i+1,2j+\frac{1}{2}} &= \left(V_{i,j+\frac{1}{2}} + V_{i,j-\frac{1}{2}}\right) / 2 \,.
	  \label{sub_ops}
\end{align}

The notations in the above equations refer to the description
of the coarse and sub-grid variables in Fig. 
The first-order interpolation applied to the coarse grid
velocity field ensures discrete incompressibility
at the sub-grid level, which is a direct consequence 
of the solenoidal nature of the coarse grid velocity field.
The choice of interpolation order used in the prolongation operator
is identical to that of the originial method of Rudman (\cite{rudman1998volume}).
The use of higher order interpolation schemes would necessitate
finding the solutions to local Poisson problems, in order to
render the sub-grid velocity field discretely divergence-free.
In the present version of our method, we have chosen to eschew the added
complexity of solving local Poisson problems
, therefore sticking with the much simpler interpolations.

\paragraph{\textbf{Restriction}}

\begin{figure}[h!]
\centering
\includegraphics[width = \textwidth]{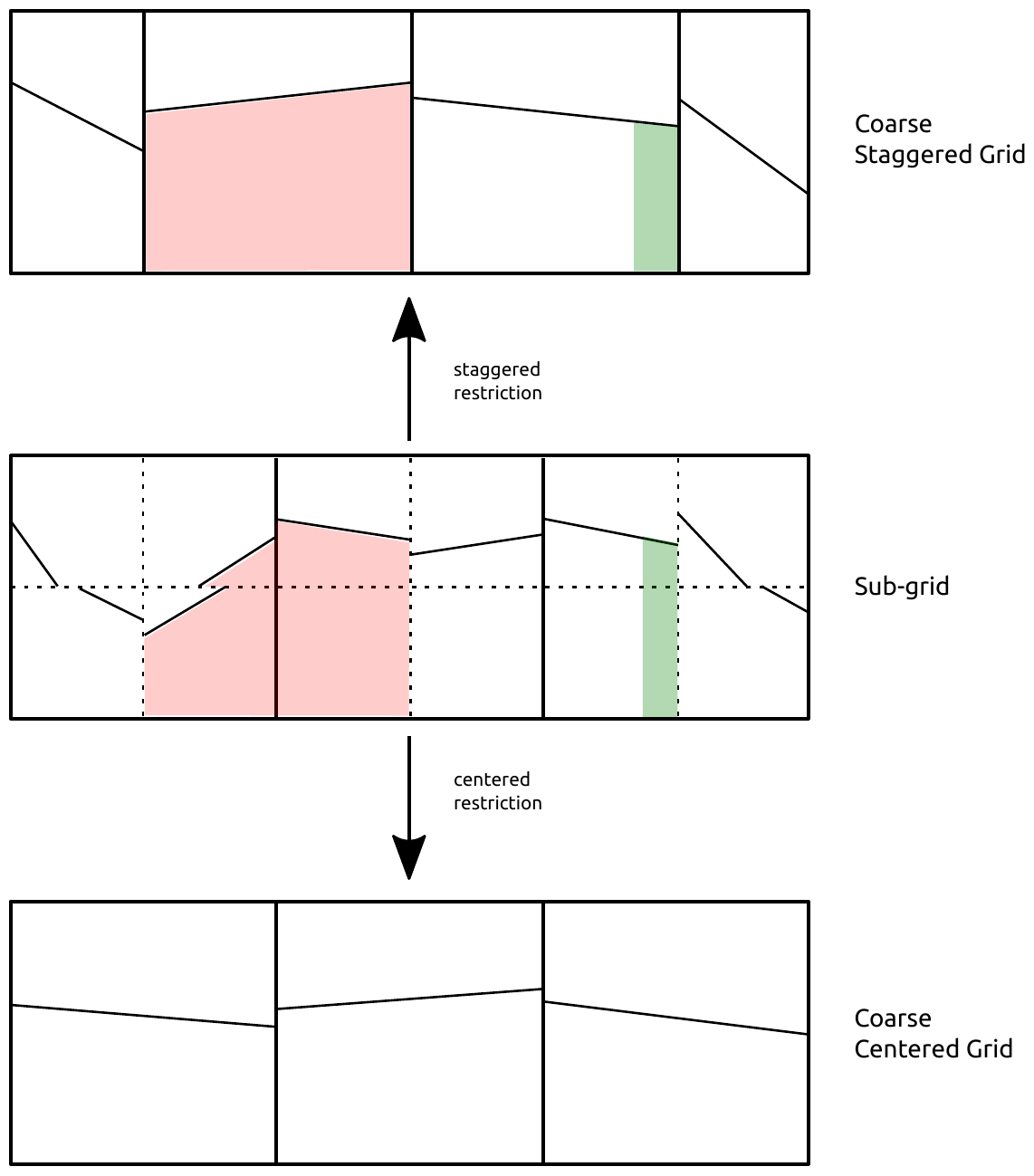}
\caption{A 2D illustration of the restriction operations involved
        between the different grid resolutions, where the sub-grid
        grid boundaries are denoted by the dashed lines,
        and that of the overlapping coarse grid by solid lines.
        Restriction operations in the form of simple volume integrals
        are performed on the sub-grid volume fraction in order
        to compute the staggered and centered volume fraction
        fields at the coarse grid level.
        Restriction operators in the form of surface integrals
        are also performed in order to compute the volume fluxes
        at the boundaries of the coarse staggered control volumes.
        The red shaded areas correspond to the volume fractions,
        whereas the green shaded area corresponds to that of volume fluxes.
        The interface segments as depicted at the coarse grid level
        are just for purpose of illustration, and are not
        `reconstructed'. 
        }
\label{restrict}
\end{figure}

In order to compute the required density and momentum fields
on the staggered grid (coarse level), we implement restriction operators
that use information from the sub-grid volume fraction field.
The restriction operators work in an identical manner as in
Rudman (\cite{rudman1998volume}), which are nothing but simple volume
integrals of the sub-grid field mapped onto the domains corresponding
to the coarse grid control volumes.
In Fig. , we demonstrate the different volume integrals for the centered and staggered fields, 
as well as the surface integrals for their corresponding fluxes.
A description of the different functions performed by the restriction
operator is as follows :

\begin{itemize}
        \item \textit{Fine Grid $\longmapsto$ Coarse Staggered Grid}

          \begin{description}
                  \item[Density \& Momentum] The sub-grid volume fraction field
                          is restricted to obtain a staggered density field at the
                          coarse grid level (see Fig. \ref{restrict}), which on combining with
                          the appropriate component of the velocity field
                          produces the staggered momentum field.
                  \item[Fluxes] The geometric fluxes from the sub-grid volume fraction
                        field are restricted in order to obtain coarse grid fluxes for the
                         corresponding staggered volumes, which on combining with
                          relation \eqref{frou} gives us the corresponding momentum fluxes.
          \end{description}

  \item \textit{Fine Grid $\longmapsto$ Coarse Centered Grid}

          \begin{description}
                  \item[Density \& Viscosity] The sub-grid volume fraction field is
                          restricted to compute a centered volume fraction field at the
                          coarse grid level, which is subsequently used to derive the
                          centered density and viscosity fields at the coarse grid level.
          \end{description}

\end{itemize}

\subsubsection{Algorithm}

To summarize the operations performed at each time step
pertaining to the advection operator in our one-fluid Navier-Stokes framework,
we present Fig. \ref{momcons_sagar}, which illustrates a 2D version
of the sub-grid method that ensures consistent and conservative
mass-momentum transport, highlighting the interactions between the
variables defined on the different grids. The algorithm is also
summarized in the following steps :

\begin{figure}[h!]
\centering
\includegraphics[width = \textwidth]{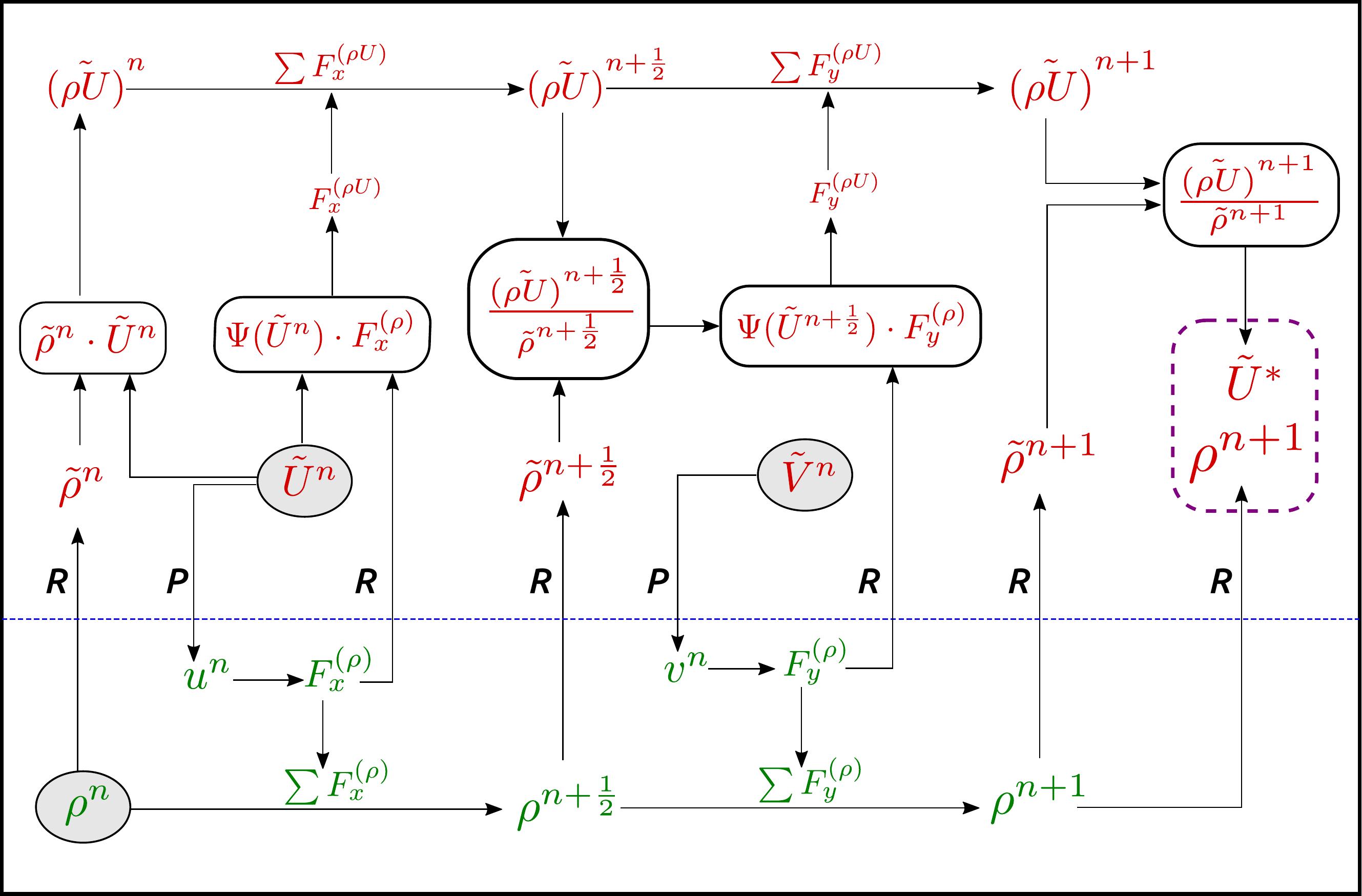}
\caption{A bird's eye view of the present method,
highlighting the operations performed at each time step.
For sake of clarity, present a 2D case for the
density and the horizontal velocity $\tilde{U}^n$
which is defined on the staggered grid $i+1/2,j$.
The variables in red are all defined on the coarse level,
and those in green are at the sub-grid level (below the blue dashed line).
The operators $R$ and $P$ denote the restriction and
prolongation operations respectively.
The ``tilde'' on top of the variables (e.g. $\tilde{\rho}^{n} $) are
used to convey that the variables are defined on the
staggered control volumes at the coarse grid level, whereas those without it  
are defined on the centered control volumes at the coarse grid level.
The operator $\Psi$ applied to the velocity field (e.g. $\Psi(\tilde{U}^n)$) 
represents the non-linear slope (flux) limiters that operate on local stencils 
, in order to estimate the momentum per unit mass
exiting the control volume during the advection substep.
The central model (\eqref{frou}) of our mass-momentum consistent method 
is represented in the operation boxes (e.g. $\Psi(\tilde{U}^n) \cdot F_x^{(\rho)}$),
thus allowing us to contruct momentum fluxes from the corresponding mass fluxes.
The starting variables are enclosed in the gray elliptical boxes,
and the variables at the end of the direction-split
integrations are $\tilde{U}^{*}$ and $\rho^{n+1}$,
where $\tilde{U}^{*}$ is subsequently fed into the RHS of the Poisson problem.
The centered density field (coarse grid) at the end of the time step ($\rho^{n+1}$)
is used by the surface tension and viscous diffusion operators.
}
\label{momcons_sagar}
\end{figure}

\begin{enumerate}

\item Interface reconstruction at time $t_n$, at the sub-grid level,
        using data $c^{n}$.

\item Restriction of the sub-grid volume fraction in order
	to compute the staggered fraction fields $C^{n}_q$ and $\rho^{n}_q$
        in the coarse grid control volumes, where $q=1,2$ is the component index.

\item Computation of all momentum components $(\rho_q u_q)^{n}$
        at $t_n$.

\item Advection of the sub-grid volume fraction $c^n$ along one coordinate direction,
        say $x$ direction, to obtain $c^{n + \frac{1}{2}}$ , using the Weymouth-Yue explicit advection method.

\item Advection of all momentum components along the $x$ direction,
        in sync with the sub-grid VOF advection,
        using \eqref{sumfmom2} to obtain the updated momentum components
                $(\rho_q u_q)^{n + \frac{1}{2}}$ after the first substep.

\item Restriction of the sub-grid field $c^{n + \frac{1}{2}}$ as shown in Fig. \ref{restrict}
        , in order to compute the ``shifted'' density field $\rho^{n + \frac{1}{2}}_q$ on the staggered volumes.

\item Extraction of the provisional velocity components. 
       $u_q^{n + \frac{1}{2}}$ after the first substep,
       i.e. $u_q^{n + \frac{1}{2}} = (\rho_q u_q)^{n + \frac{1}{2}}/\rho_q^{n + \frac{1}{2}}$.
	These provisional velocities are required to compute the 
	``advected'' velocities using the non-linear flux limiters. 

\item The above operations are repeated for the different momentum components,
        in sync with the direction-split advection of the sub-grid volume fraction
        along the remaining direction.
        At each time step, the sequence $x, y$ is permuted, in order to avoid
        any systematic biases in error propagation.
        Finally, we obtain $u_q^{*} = (\rho_q u_q)^{n + 1}/\rho_q^{n + 1}$ on the staggered grid.

\item In parallel, the updated centered volume fraction field $C^{n+1}$ is obtained
        at the coase grid level, by applying the restriction operator
        on the updated sub-grid volume fraction $c^{n+1}$.

\end{enumerate}

The resulting method not only maintains discrete consistency between
mass and momentum transport on the staggered grid, but also ensures that
the transport of momentum is conservative in 3D.
The other time-split terms in equation (\ref{mom_update})
that arise from the source term discretizations, as well
as the terms in the projection step \eqref{ufinal} are solved in
a standard non-conservative way, and are briefly covered in what follows.
The density on the faces of the central cells
are estimated using the restriction operations
described previously, with these densities appearing
via the $ 1 / \rho_q $ pre-factor in front of all
the the other terms involved in momentum transport (e.g surface tension, viscous diffusion).

\subsection{Surface Tension and Viscous Diffusion}
The detailed descriptions of the methods used in our
numerical platform PARIS Simulator \cite{paris} to deal with surface tension,
viscosity and body forces have already been carried
out in \cite{paris, basilisk, popinet2009accurate},
therefore we only briefly touch upon certain aspects of the
operators in question, in particular,
their interaction with the volume fraction field.

\paragraph{Surface Tension}
We use the Continuum Surface Force method (CSF) as our model for surface tension,
coupled with height functions for curvature computation.
The height functions used in our implementation were first introduced in \cite{popinet2009accurate}
, subsequently tested, revised and improved in \cite{bornia2011properties,owkes2015mesh}.
In general, the height functions are used to compute the curvature field based
on second-order finite differences applied to the heights.
The height functions are derived from the centered volume fraction field 
at the coarse grid level, which itself is computed via the previously discussed restriction operations.
Although, in regions of poor interfacial resolution where the local radius of curvature 
is comparable to the grid size, the method reverts to certain fallbacks such as curve fitting.
The resulting curvature field is coupled with a well-balanced discretization with
respect to the discrete pressure gradient, with the same discretization
stencil applied to the volumetric (body) force term as well.
Interestingly, the sub-grid volume fraction can be used to get better estimates 
of the curvature field, whose estimates are required on staggered control volumes.
At the time of writing, two different approaches are being
developed and tested in order to get more accurate curvature fields, 
but in the context of the present study we refrain from any further complexity.

\paragraph{Viscous Diffusion}
We use second-order spatial discretizations of the viscous stresses, using centered differences.
The (variable) dynamic viscosity computations are based on the
centered volume fraction field at the coarse grid level, as expressed in \eqref{mu_chi}.
The temporal treatment of the viscous term can be either in explicit or semi-implicit fashion,
, but in the context of the present study we will be sticking exclusively with the explicit version.

\subsection{Pressure-Poisson Projection}
The velocity field is evolved using a classical time-splitting projection method
as described in the seminal work of Chorin \cite{chorin1969convergence},
which involves predicting an ``intermediate'' velocity field $\boldsymbol{u}^{*}$
as given by equation \eqref{mom_update}, followed by a correction step given by 

\begin{align}
\boldsymbol{u}^{n+1} = \boldsymbol{u}^{*} - \frac{\tau}{\rho^{n+1}}\nabla^{h}p^{n+1} \,.
\label{ufinal}
\end{align}

The discrete pressure field required to correct the intermediate velocity
is determined by imposing the conservation of mass, which in our incompressible
framework reduces to necessitating the resulting velocity field to be divergence-free (solenoidal)

\begin{align}
\nabla^{h}\cdot\boldsymbol{u}^{n+1} = 0 \,.
\label{div}
\end{align}

Thus, combining equations \eqref{ufinal} and \eqref{div},
we are left with a variable coefficient Poisson equation for the pressure, given as  

\begin{align}
\nabla^{h}\cdot\left(\frac{\tau}{\rho^{n+1}}\nabla^{h}p^{n+1}\right) = \nabla^{h}\cdot \boldsymbol{u}^{*} \,.
        \label{poisson}
\end{align}

The Poisson solver used in PARIS Simulator to invert the elliptic operator
appearing in eqn. \eqref{poisson} is a red-black
Gauss-Seidel (GS) solver with overrelaxation \cite{briggs1987}.
There is also has an in-house implementation of a multigrid solver
for structured grids with $2^{n}$ number of points per direction,
utilizing a fully parallelized V-Cycle scheme \cite{briggs1987}.
Relaxation operations are applied starting from the finest to the coarsest first,
and then from the coarsest to the finest, with the number of relaxation operations
being a user-adjustable parameter. Having a native multigrid solver allows
for an efficient solution of the Poisson equation without the necessity
of having external libraries/pre-conditioners (e.g. HYPRE) installed on the system.


\section{Numerical Benchmarks}
\label{sec:bench}
In this section we carry out quantitative comparisons between the solutions 
obtained by our present mass-momentum consistent and conservative method, 
and certain analytical, semi-analytical, equilibrium or asympotic solutions 
which correspond to well-established benchmark configurations.  
Generally, these established test cases in existing literature (e.g. decay of spurious 
currents in static and moving droplets, viscous damping of capillary waves etc.), 
are carried out in the absence of any density jump (or viscosity jump) across the interface.
Thus, we are distinctly interested in the behavior of our present method 
as it deals with the non-linear coupling between interfacial deformation/propagation, 
capillary and viscous forces, especially in regimes where fluid densities  
are separated by orders of magnitude across the interface. 
In the interest of clarity, we shall adopt the following nomenclature from this point onwards 

\begin{itemize}
	\item \textbf{STD} : Standard version of our method based on the non-conservative formulation of Navier-Stokes equations,
does not maintain discrete consistency between mass and momentum transport. 
	\item \textbf{MSUB} : Present method which ensures consistent and conservative transport of mass-momentum.  
\end{itemize}

\subsection{Static Droplet} 
\label{subsec:sdrop}
A popular numerical benchmark in the existing literature relevant to surface 
tension dominated flows is the case of a spherical droplet of a denser 
fluid immersed in a quiescent surrounding medium of lighter fluid. 
In the hydrostatic limit of the Navier-Stokes equations, 
the droplet should stay in equilibrium, with a curvature induced pressure jump 
across the interface corresponding to Laplace's equilibrium condition. 
In practice however, numerically reproducing such a trivial equilibrium 
condition is not as straighforward, as there exists a slight difference between 
the initial numerical interface and the exact analytical shape of the sphere, 
thereby resulting in the generation of the well documented '\textit{spurious}' 
or '\textit{parasitic}' currents of varying intensity in the velocity 
field \cite{lafaurie1994modelling, harvie2006analysis, popinet1999front}. 
Considerable progress has been made by the adoption of \textit{well-balanced} 
surface tension formulations, that ensure consistency between the numerical 
stencils used for the discretization of the pressure gradient and 
$n \delta_{s}$ (\cite{francois2006balanced, popinet2009accurate}). 
In this context, Popinet \cite{popinet2009accurate} demonstrated that 
given sufficient time (order of viscous dissipation time scales), 
a well-balanced method will relax to the '\textit{numerical}' equilibrium 
shape through damping of the 'physically consistent' numerical capillary waves, 
therefore allowing us to recover the exact (to machine precision) Laplace equilibrium condition.   


\begin{figure}[h!]
    \centering
    \includegraphics[scale=0.5]{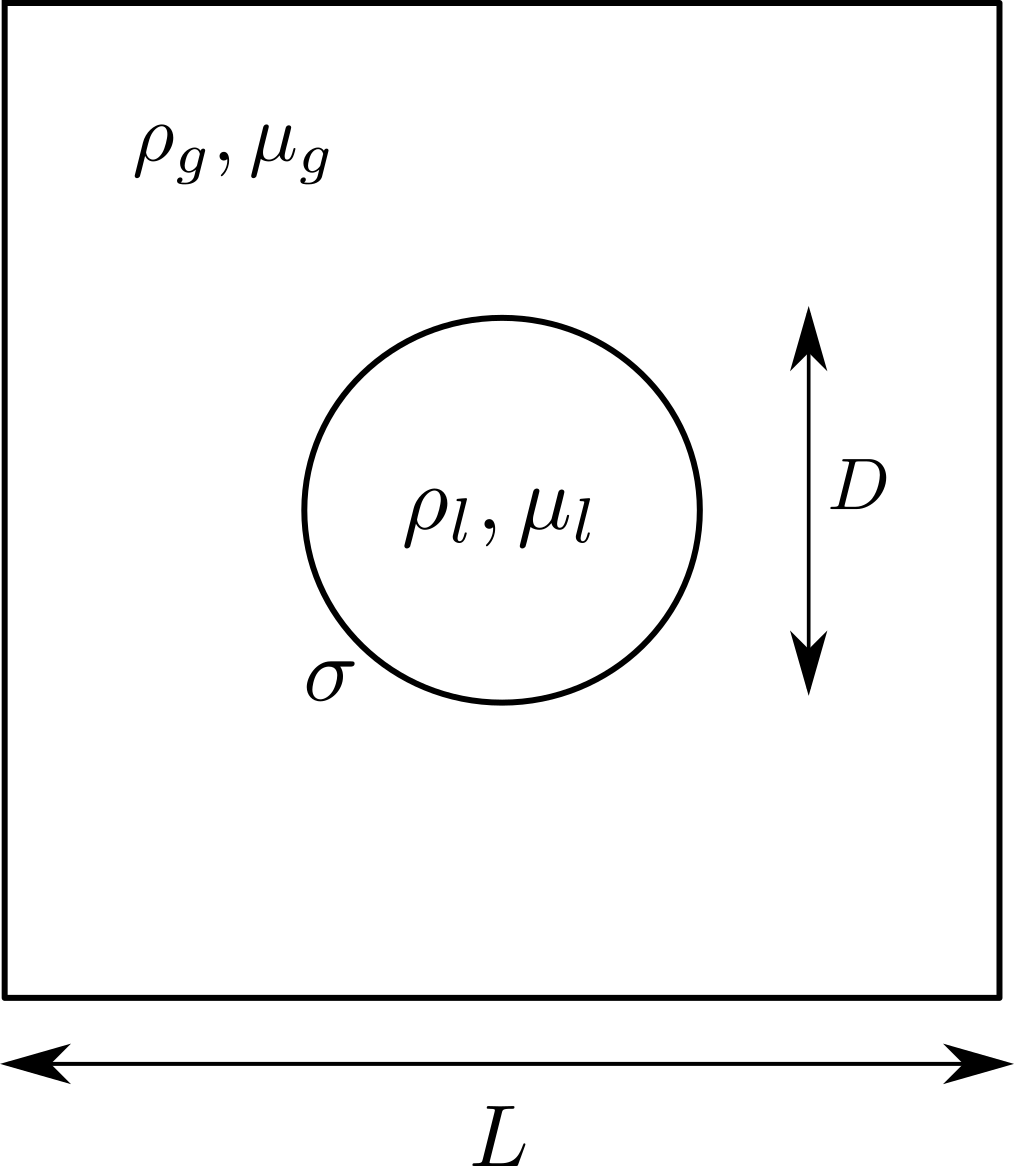}
    \caption{Schematic of the static droplet of dense fluid surrounded by a quiescent medium of lighter fluid. A $40 \times 40$ grid is employed to spatially discretize the domain.} 
    \label{static_conf}
\end{figure}

The key difference in our implementation of this classic test case from 
that of Popinet \cite{popinet2009accurate} is that we consider the effect of 
density contrast across the interface separating the fluids. 
A sharp density jump across the interface may exacerbate the effect of 
numerical errors incurred as a result of interfacial reconstructions, 
curvature estimation and various other truncations. 
We consider a circular droplet of size $D$  placed at the centre 
of a square domain of side $L$. 
The densities of the heavier and lighter phases are $\rho_l$ and $\rho_g$ respectively, 
likewise for the viscosities $\mu_l$ and $\mu_g$, and $\sigma$ 
being the surface tension coefficient (fig. \ref{static_conf}). 
The ratio of the droplet size to the box is chosen as $D/L = 0.4$,
and the numerical resolution is $D/\Delta x= 16$ ($\Delta x$ is the grid size). 
We use symmetry conditions on all sides of our square domain.        
The problem incorporates two natural time-scales, the capillary oscillation 
scale and the viscous dissipation scale, which are defined as  

\begin{align}
	T_\sigma = \left(\frac{\rho_l D^3}{\sigma}\right)^{1/2} \quad , \quad T_\mu = \frac{\rho_l D^2}{\mu_l} \,. 
\label{ts}
\end{align} 

The ratio of these time-scales give us  

\begin{align}
	\frac{T_\mu}{T_\sigma} = \sqrt{\rho_l \sigma D}/\mu_l = \sqrt{\textrm{La}} \,,
\end{align}

where $\textrm{La}$ is the Laplace number based upon the heavier fluid. 
In the present study, we introduce the density-ratio $\rho_l/\rho_g$ as another important parameter. 
We rescale our ''parasitic'' velocity field using a velocity scale as defined below 

\begin{align}
	U_\sigma = \sqrt{\sigma/\rho_l D} \,. 
\end{align}

\begin{figure}[h!]
    \centering
	\includegraphics[width = 1.0\textwidth]{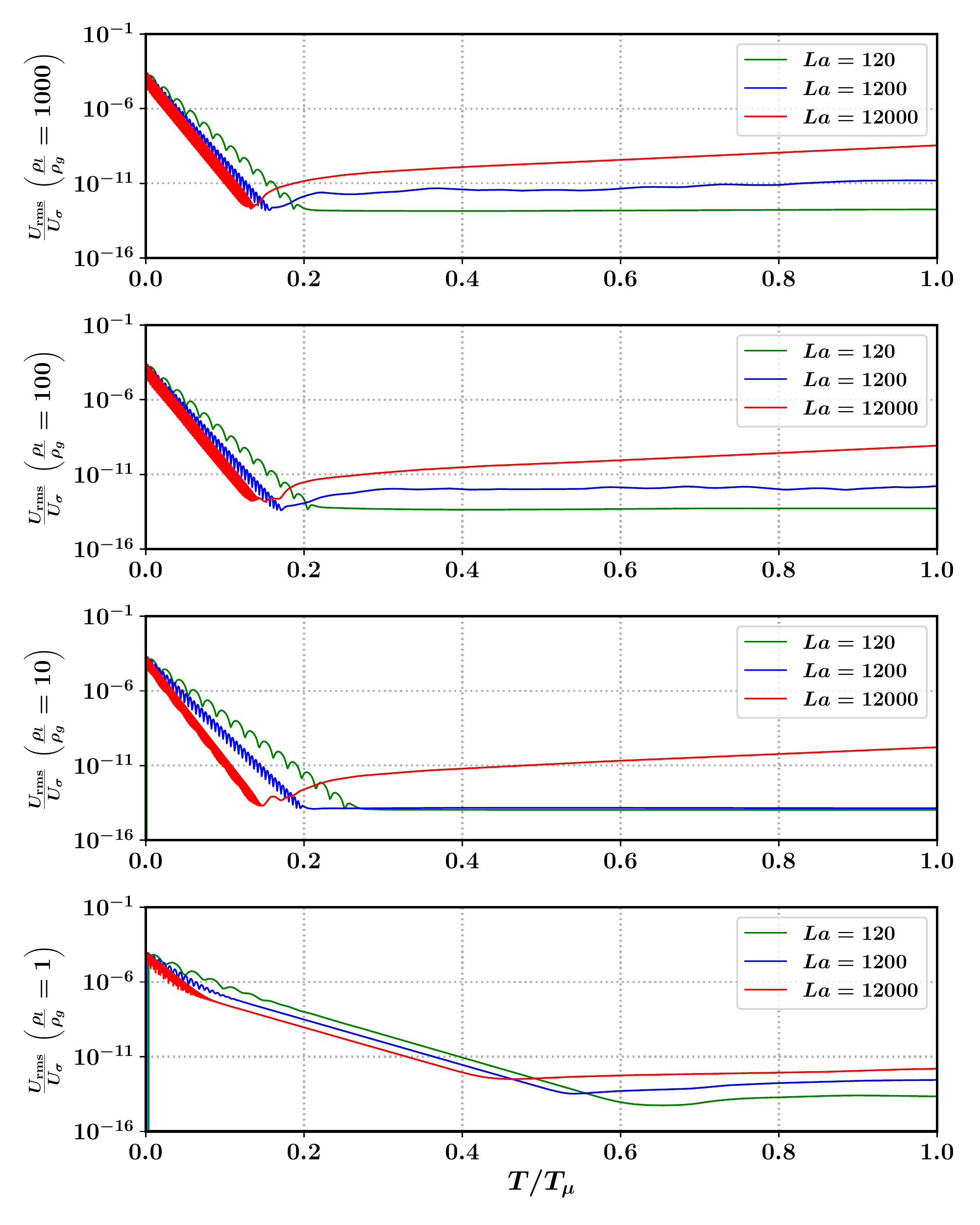}
	\caption{\textbf{STD} : Decay of normalized spurious currents as a function of viscous dissipation time-scales for different density-ratios and Laplace numbers. The currents seem to initially decay quickly for larger density-ratios, and relax to the numerical equilibrium curvature within $0.2 \cdot T_\mu$. For combinations of large $\rho_l / \rho_g$ and large $\textrm{La}$, the spurious currents seem to grow back to an order of magnitude ($10^{-8}$) which is quite far from that of machine precision ($10^{-14}$).}   
    \label{decay_nonmc}
\end{figure}

\begin{figure}[h!]
    \centering
	\includegraphics[width = 1.0\textwidth]{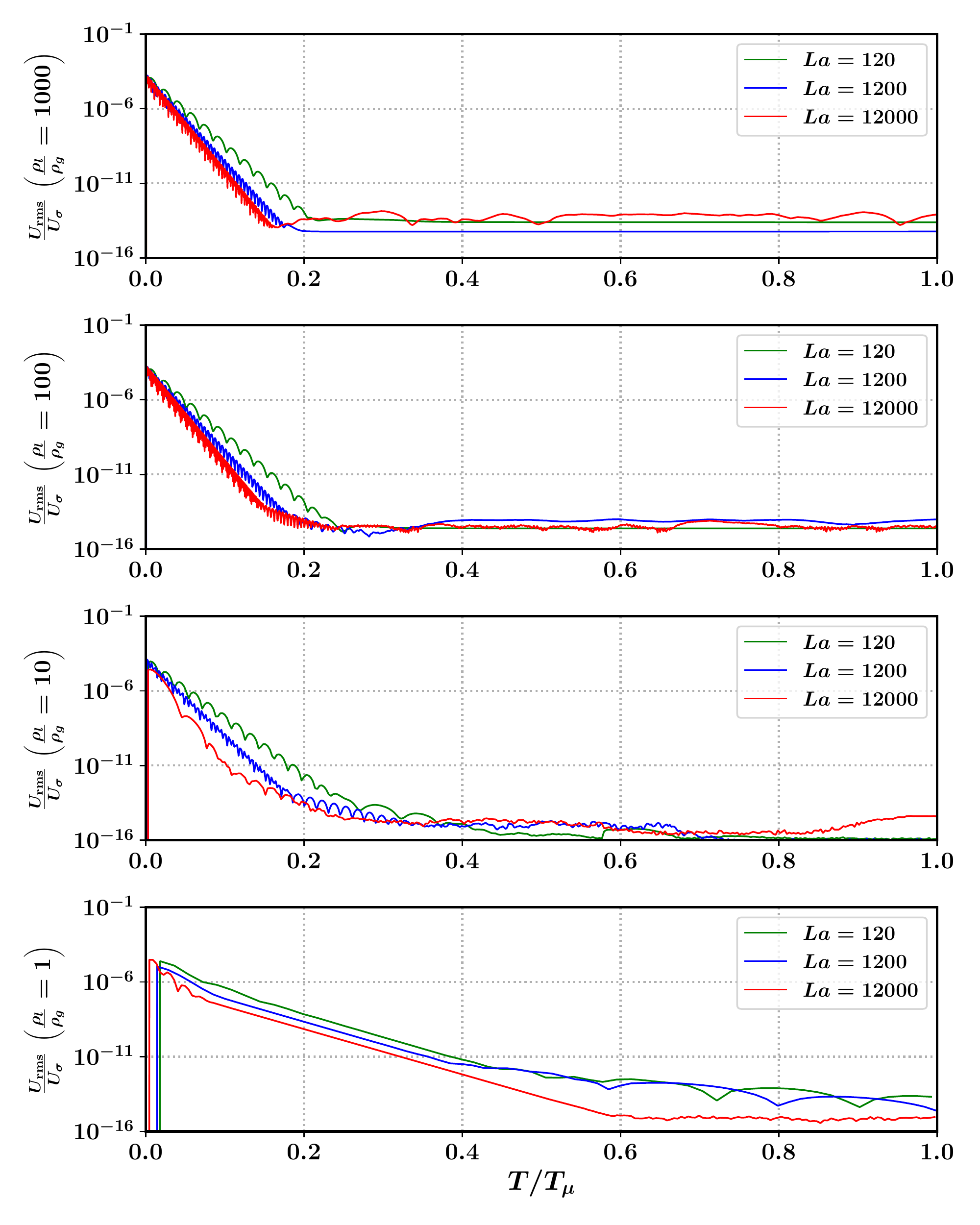}
	\caption{\textbf{MSUB} : Decay of normalized spurious currents as a function of viscous dissipation time-scales for different density-ratios and Laplace numbers. The currents seem to decay very quickly in the case of large density-ratios, and relax to the numerical equilibrium curvature within $0.2 \cdot T_\mu$. For all combinations of $\rho_l / \rho_g$ and $\textrm{La}$ numbers, the decayed spurious currents are not observed to grow back as in the cases of \textbf{STD}, and hover around values close to machine precision ($10^{-14}$).}   
\label{decay_sagar}
\end{figure}

Additionally, due to the explicit nature of our surface tension model,
the time-step in our numerical simulation must be smaller than 
the oscillation period corresponding to the grid wavenumber 
(fastest capillary wave $\sim \left( \rho_l \Delta x^3 / \sigma  \right)^{1/2} $ ). 
For the scope of the present study, we shall not consider any viscosity contrast between 
the two fluids while varying the density contrast, therefore setting $\mu_l/\mu_g = 1$ for all the cases.

\begin{figure}[h!]
    \centering
    \includegraphics[width = 0.9\textwidth]{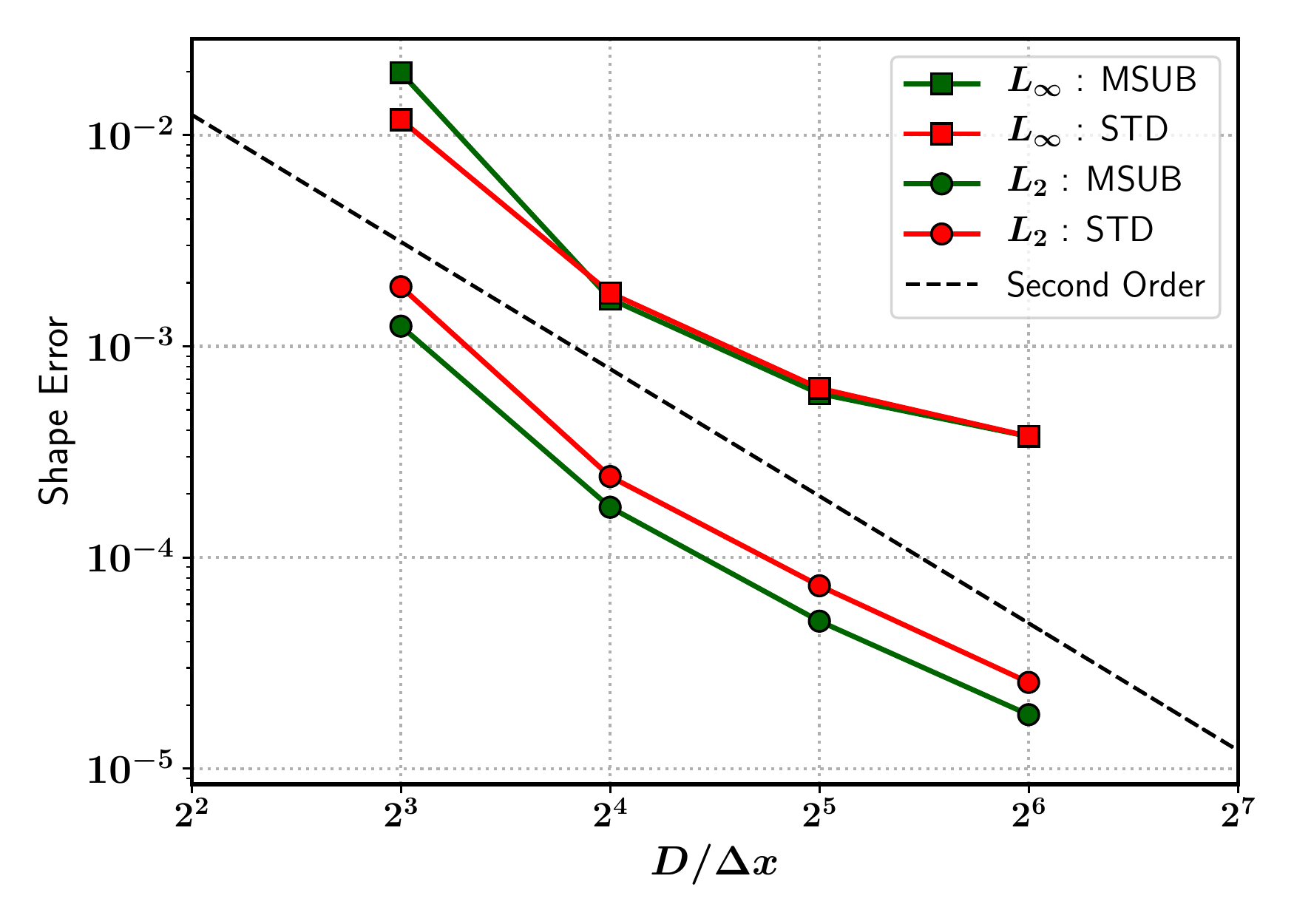}
	\caption{Second-order spatial convergence for the spurious current error norms corresponding to a stringent parameter combination ($\rho_l/\rho_g = 1000$ , $\textrm{La} = 12000$) . Both of the norms ($L_\infty$ and $L_2$) seem to demonstrate a roughly second order rate of spatial convergence with each of the methods tested. However, the $L_2$ errors of \textbf{MSUB} are smaller by approximately a factor of $2$ compared to \textbf{STD} for all resolutions tested.}   
    \label{static_conv}
\end{figure}

In figures \ref{decay_nonmc} and \ref{decay_sagar}, we illustrate the decay of the 
root-mean-square of the spurious currents as a function of time, 
in the case of four different density contrasts, 
and with three different Laplace numbers for each ratio. 
Figure \ref{decay_nonmc} refers to simulations carried out with the non-consistent (\textbf{STD}) method,
and Fig. \ref{decay_sagar} corresponds to that using the present method (\textbf{MSUB}). 
Time is rescaled by the viscous dissipation scale, and the spurious currents by the capillary velocity scale. 

We have two main observations, first being the rapid decay of the rescaled spurious currents 
for all combinations of density contrasts and Laplace numbers within approximately $0.2 T_{\mu}$. 
Secondly, there is a slower re-growth of the currents for combinations of non-unity
density contrasts and large Laplace numbers, 
for all simulations except those carried out with the present method (\textbf{MSUB})
With the present method, the once decayed currents keep hovering around 
levels of machine precision for remainder of time. 
Therefore, the present method demonstrates the desired performance, 
systematically for all combinations of large density contrasts 
coupled with large Laplace numbers. 

Once the solution relaxes to a numerical equilibrium curvature,
there still exists a difference between the numerical curvature and 
the exact analytical curvature corresponding to the spherical (circular) shape. 
We use the definitions of the shape errors as introduced in 
the seminal work of Popinet \cite{popinet2009accurate} to assess 
the convergence to the exact (analytical) curvature as we increase spatial resolution. 
The norms are defined as      

\begin{align}
	L_2 = \sqrt{\frac{\sum_i \left(C_i - C_i^\textrm{exact} \right)^2}{\sum_i}} \quad , \quad L_\infty = \textrm{max}_i \left( | C_i - C_i^\textrm{exact} | \right) \,,
  \label{shape_err}
\end{align}

where $C_i$ is the volume fraction of a cell after the solution has 
relaxed to the numerical equilibrium curvature, and $C_i^\textrm{exact}$ 
is the volume fraction corresponding to the exact circular shape.  
The fields $C_i$ and $C_i^\textrm{exact}$ correspond to the coarse grid level,
and are obtained via restriction of the sub-grid volume fraction field. 
Fig. \ref{static_conv} demonstrates the behavior of the shape errors defined 
in \eqref{shape_err} for the case of the most stringent parameter combination 
( $\rho_l / \rho_g = 1000 $ , $\textrm{La} = 12000$ ), as a function of the droplet resolution. 
One can clearly observe that both methods display a roughly second-order convergence in space for the error norms. 
In terms of the $L_2$ norm, the present method (\textbf{MSUB}) achieves smaller errors (roughly by a factor of 2) 
as compared to the non-consistent method (\textbf{STD}) for all resolutions tested.

\subsection{Droplet Advection}
\label{subsec:mdrop}
An incisive numerical setup which evaluates the accuracy of the coupling between 
interfacial propagation and surface tension discretization was first proposed by 
Popinet \cite{popinet2009accurate}, subsequently employed in 
the comparative study of Abadie et al. \cite{abadie2015combined}. 
This test differs from that of the static droplet due to the presence 
of a uniform background velocity field, therefore serving as a better 
representation of droplets in realistic surface tension dominated flows 
where they might be advected by the mean flow.        

\begin{figure}[h!]
    \centering
    \includegraphics[scale=0.5]{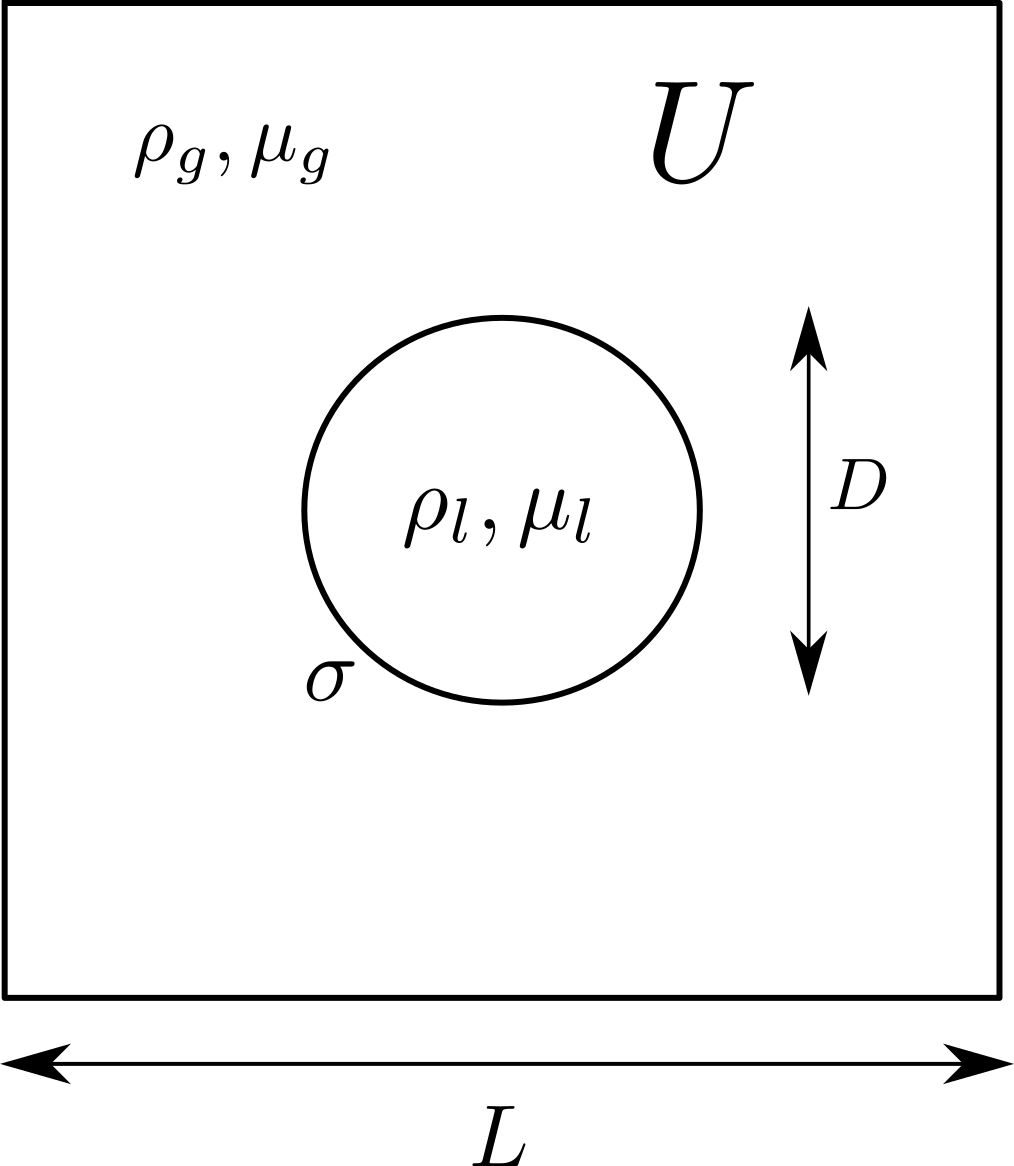}
    \caption{Schematic of the droplet of dense fluid advected in a surrounding medium of lighter fluid. A $50 \times 50$ grid is employed to spatially discretize the domain, which is spatially periodic in the direction of droplet advection.} 
    \label{moving_conf}
\end{figure}

In terms of the Laplace equilibrium, the hydrostatic solution is still valid 
in the frame of reference of the moving droplet. 
The solution in the moving reference frame diverges from that of the 
static droplet as a result of the continuous injection of noise at the grid scale. 
This noise originates from the perturbations to the curvature estimates, 
which are themselves born out of the interfacial reconstructions needed to advect the interface. 
This transforms the problem into that of viscous dissipation with continuous forcing (in the moving reference frame).   

\begin{figure}[h!]
    \centering
    \includegraphics[width = 1.0\textwidth]{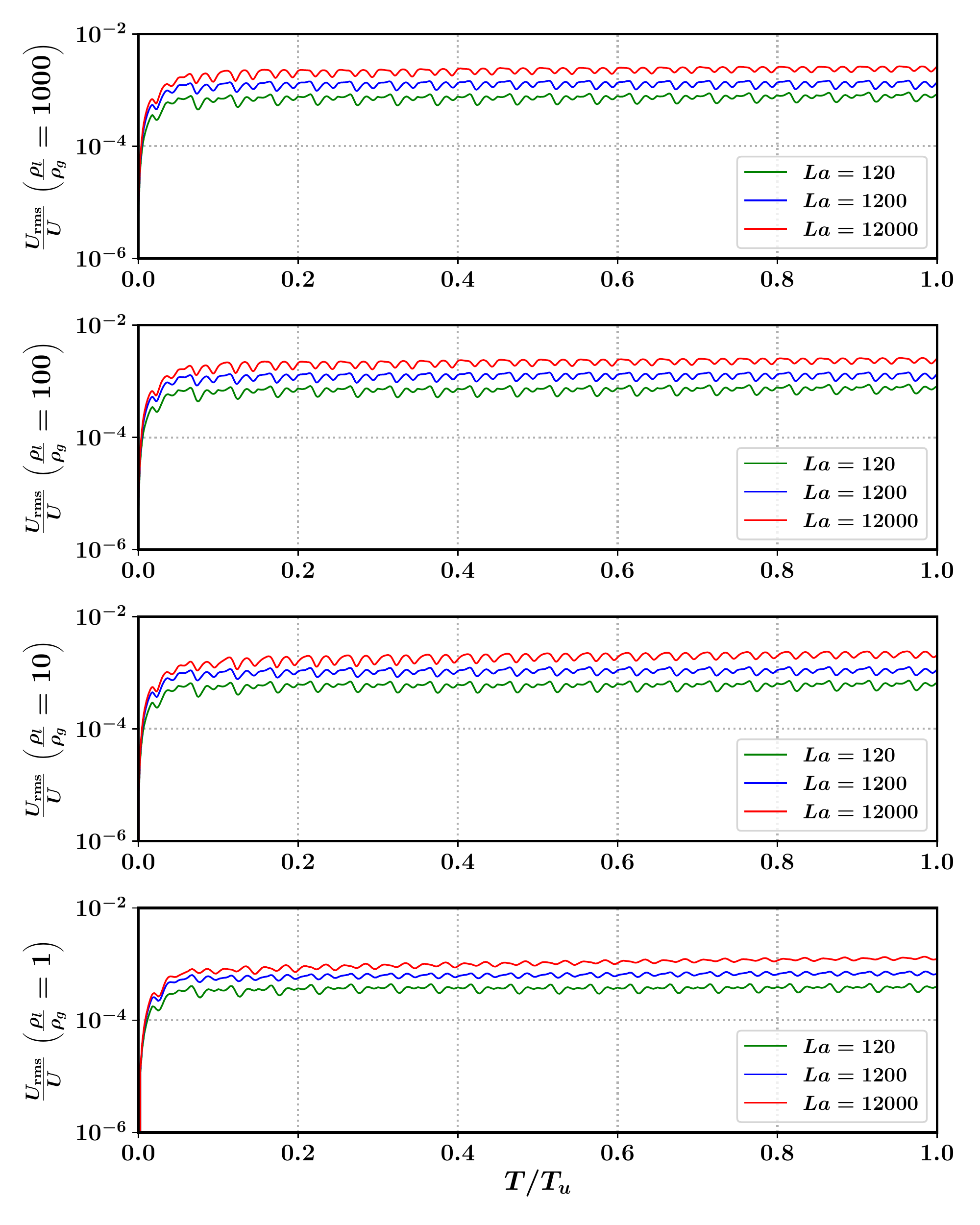}
	\caption{\textbf{STD} : Time evolution of normalized spurious currents as a function of advection time-scales ($T_u$) for different combinations of density-ratio and Laplace numbers. The currents seem to hover around $10^{-3}$, with a larger Laplace number corresponding to a higher error for all density-ratios. $\textrm{We} = 0.4$ for all the cases presented.}   
    \label{evo_nonmc}
\end{figure}

In the present study, we evaluate our present method using the advection of a 
droplet in a spatially periodic domain in the same setup as \cite{popinet2009accurate}, 
but with the important difference of including sharp density jumps across the interface. 

\begin{figure}[h!]
    \centering
    \includegraphics[width = 1.0\textwidth]{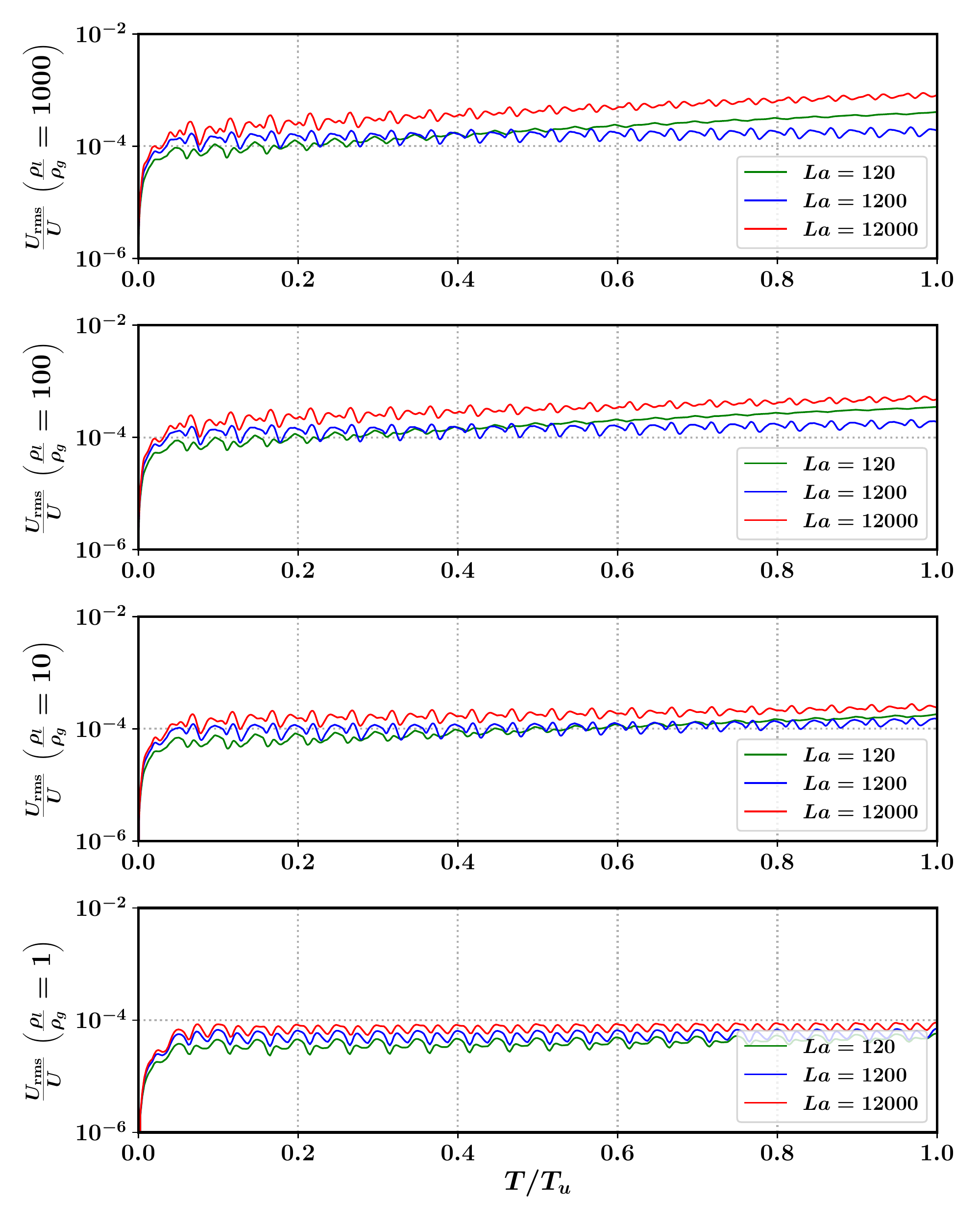}
	\caption{\textbf{MSUB} : Time evolution of normalized spurious currents as a function of advection time-scales ($T_u$) for different combinations of density-ratio and Laplace numbers. In terms of the errors observed in \textbf{STD}, we observe a decrease of roughly one order of magnitude. Although an upward trend is observed for large Laplace numbers, the growth rate is quite low. The currents seem to hover slightly above $10^{-4}$, with larger Laplace numbers corresponding to larger errors for all density-ratios. $\textrm{We} = 0.4$ for all the cases presented.}   
    \label{evo_sagar}
\end{figure}

We consider a circular droplet of diameter $D$ placed at the centre of a square domain of side $L$. 
The densities of the heavier and lighter phases are $\rho_l$ and $\rho_g$ respectively, 
likewise for the viscosities $\mu_l$ and $\mu_g$, and $\sigma$ being the 
surface tension coefficient (fig. \ref{moving_conf}). 
A uniform (horizontal) velocity field $\boldsymbol{U}$ is initialized on the entire domain. 
The ratio of the droplet size to the box is $D/L = 0.4$, with $D/\Delta x= 20$. 
For reference, Popinet \cite{popinet2009accurate} used a resolution of $D / \Delta x = 25.6$ 
corresponding to a grid of $64 \times 64$. 
We use symmetry conditions on the top and bottom sides, and periodic boundary 
conditions along the horizontal direction. 
The problem is characterized by adimensional parameters based on the heavier fluid
, given by  

\begin{align}
	\textrm{La} = \frac{\rho_l \sigma D}{\mu_l^2} \quad , \quad \textrm{We} = \frac{\rho_l U^2 D}{\sigma} \,. 
\end{align}

\begin{figure}[h!]
\centering
\begin{subfigure}{.5\textwidth}
    \centering
    \includegraphics[width = 1.\linewidth]{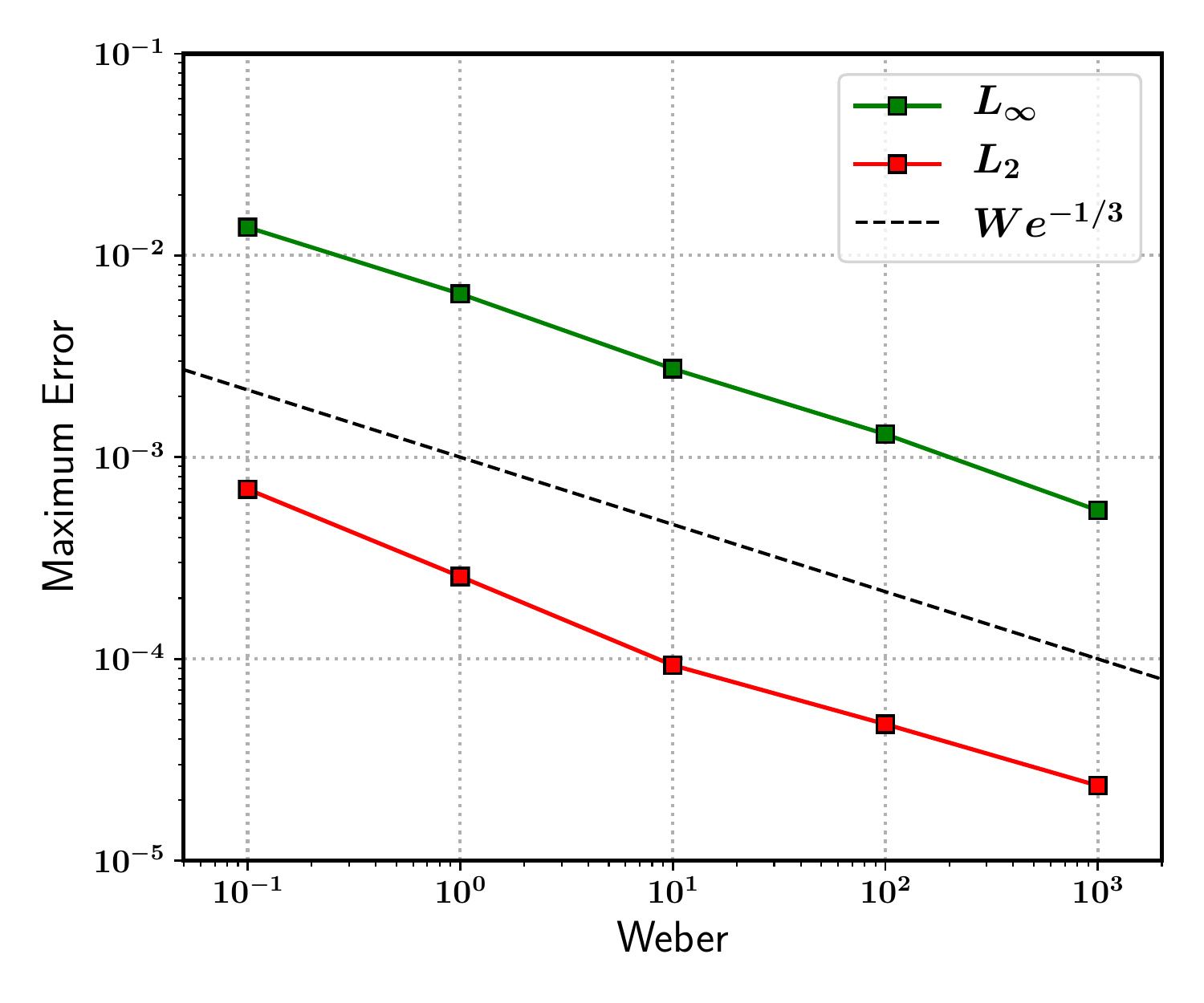}
    \label{web}
\end{subfigure}%
\begin{subfigure}{.5\textwidth}
    \centering
    \includegraphics[width = 1.\linewidth]{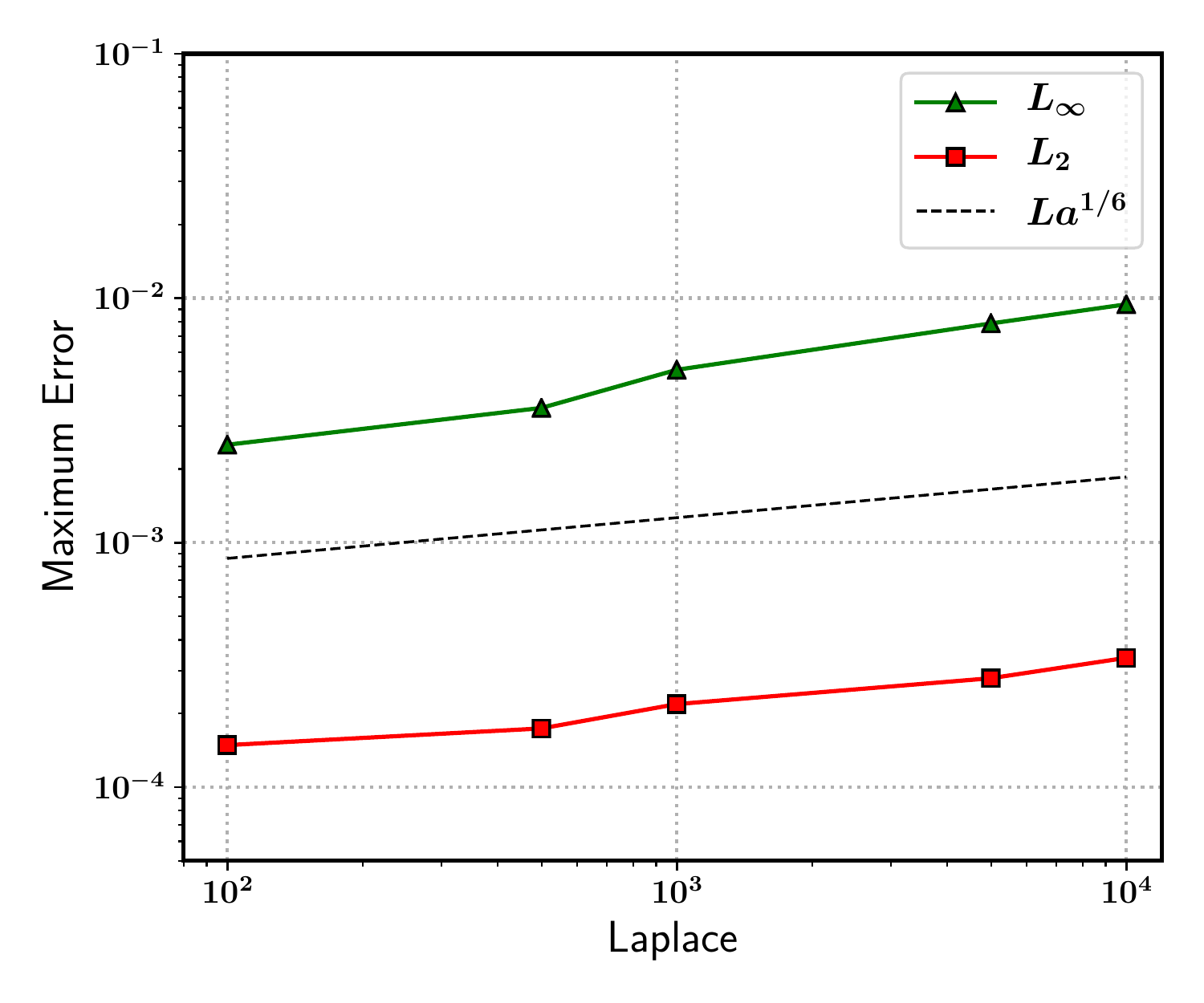}
    \label{lap}
\end{subfigure}
\caption{
	Maximum of the error norms as a function of different $We$ and $\textrm{La}$ numbers
for our present method (\textbf{MSUB}), $\rho_l/\rho_g = 1000$ for all the cases presented. 
	(a) Maximum error norm as a function of Weber ($\textrm{La} = 12000$, $\rho_l / \rho_g = 1000$).
	(b) Maximum error norm as a function of Laplace ($\textrm{We} = 0.4$, $\rho_l / \rho_g = 1000$).  
 }
\label{we-la}
\end{figure}

In addition to the capillary and viscous time-scales for the static 
case (eqns. \ref{ts}), we have an additional scale defined as  

\begin{align}
     T_U = D/U \,,
\end{align}

which is the time-scale of advection. In our subsequent analysis, 
we shall use $T_u$ and $U$ as the time and velocity scales, repectively.
Figures \ref{evo_nonmc} (\textbf{STD} method ) and \ref{evo_sagar} (\textbf{MSUB}) depict the evolution of the 
root-mean-square (RMS) error of the velocity field in the moving frame of reference, 
as a function of different Laplace numbers, spanning over several density contrasts. 
We again have a couple of important observations, the first being that spurious 
currents do not decay to machine precision as in static droplet case for 
all combinations and methods tested, instead they oscillate around a 
mean value of the order of $0.1-0.01 \% $ of the constant field $U$. 
The second observation is regarding the significantly smaller error 
(roughly by one order of magnitude) in the case of the present method (\textbf{MSUB}) 
when compared to the non-consistent one (\textbf{STD}). 
As a minor remark, in case of large Laplace numbers, the \textbf{MSUB} method displays 
a slight upward trend in the error evolution, which is not the case in \textbf{STD}. 
This is not too worrisome as the growth is over a time-scale much larger than $T_u$, 
with the faster oscillations corresponding to a time-scale of the order $U/\Delta x$. 
All of the plots correspond to $\textrm{We} = 0.4$, coupled with an additional simplification of 
having equal viscosities across the interface i.e $\mu_l/\mu_g = 1$ .
As evindenced by the persistence of these spurious currents due to the 
addition of grid-level noise emanating from interfacial reconstructions, 
further advancements should be made with respect to the combined performace
of the interfacial transport, curvature computation and the surface tension model. 

\begin{figure}[h!]
    \centering
    \includegraphics[width = 0.8\textwidth]{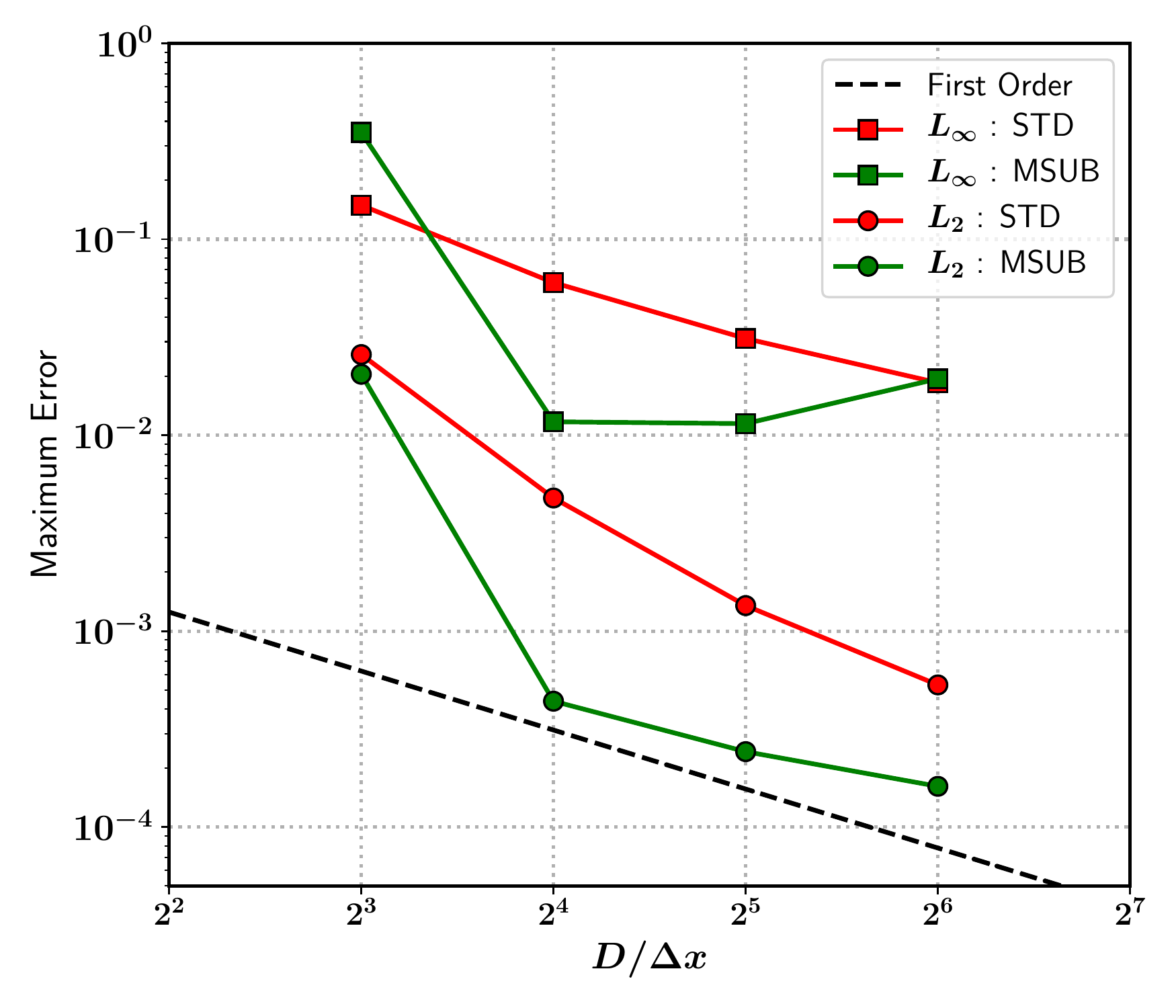}
	\caption{First-order (approximately) spatial convergence of the maximum of the spurious current error norms in the frame of reference of the moving droplet, for the most stringent parameter combination ($\rho_l/\rho_g = 1000 $ , $\textrm{La} = 12000$, $\textrm{We} = 0.4$). The consistent and conservative method (\textbf{MSUB}) leads to significantly lower errors. }   
    \label{moving_conv}
\end{figure}

In order to evaluate the performance of our methods at different resolutions, 
we define the errors as the maximum values of the norms $L_\infty$ and 
$L_2$ of the rescaled field $U_{rms}/U$ over time ($5$ times $T_u$). 
In Fig. \ref{moving_conv}, we show the scaling of the error as a function of 
spatial resolution for the most stringent case of $\rho_l/\rho_g = 1000 $ , $\textrm{La} = 12000$.
For our present method (\textbf{MSUB}), we observe significantly lower maximum errors 
compared to the non-consistent method (\textbf{STD}), but at a cost of slightly 
less than first-order convergence rate. 
The overall convergence behavior of the methods we have tested seem to be 
consistent with earlier studies of Popinet \cite{popinet2009accurate}, 
although in that study only equal densites across the interface were considered.  

As the final point of inquiry, Fig. \ref{we-la} shows the influence of the Laplace 
and Weber numbers on the behavior of the maximum error norm, 
carried out for the largest density contrast ($\rho_l/\rho_g = 1000$).
We only present the results obtained using the present method (\textbf{MSUB}), 
for a resolution corresponding to $D / \Delta x = 25.6$. 
We observe that the error (both $L_\infty$ and $L_2$) scales as 
$We^{-1/3}$ over 4 orders of magnitude, which is different from the $\textrm{We}^{-1/2}$ 
scaling observed by Popinet \cite{popinet2009accurate} (for equal densities). 
In case of Laplace numbers, the errors scale as $\textrm{La}^{1/6}$ over two orders of magnitude, 
which is identical to the that in \cite{popinet2009accurate} (again, for equal densities).

\subsection{Capillary Wave}
\label{subsec:capwave}
One of fundamental features of surface tension dominated flows are 
the propagation of capillary waves. 
A numerical method should not only be able to adequately resolve, 
but also accurately emulate the spatio-temporal evolution of such capillary oscillations. 
A brief outline on the state-of-the-art numerical implementations of capillary waves 
(and surface tension models in general) is provided by Popinet \cite{popinet2018numerical}.

\begin{figure}[h!]
    \centering
    \includegraphics[scale=0.7]{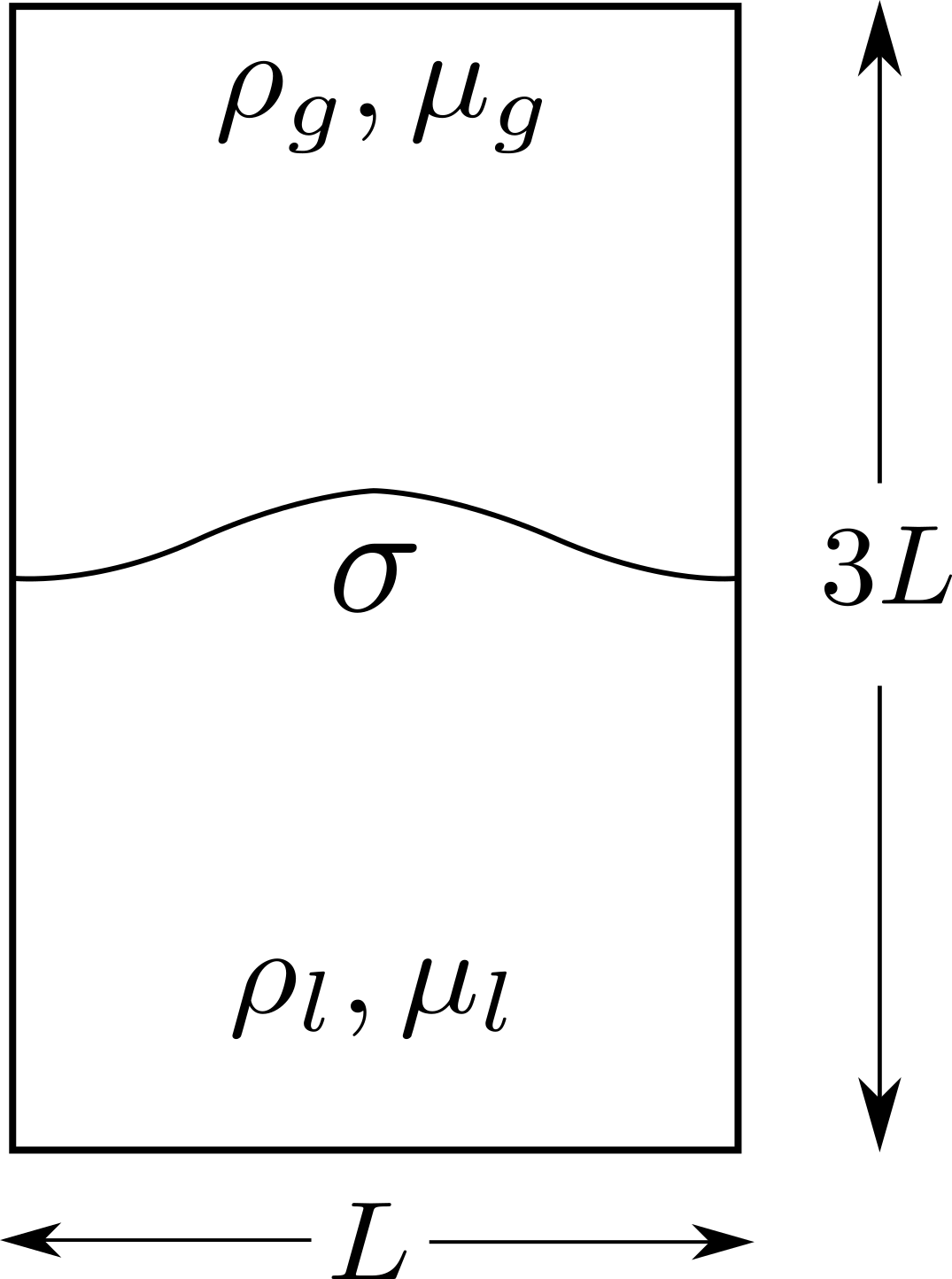}
	\caption{Schematic of the initially perturbed planar interface separating two immiscible fluids of different densities and viscosities. A spatial resolution of $32 \times 96$ is used for spatial discretization (compared to $64 \times 192$ in Popinet \cite{popinet2009accurate}), with the width of the box corresponding to the size of the perturbed wavelength.}
    \label{capwave_conf}
\end{figure}

In the present study, we evaluate the accuracy of our method by comparing 
with an analytical solution of damped capillary oscillations. 
Generally, such solutions exist only for extremely small initial perturbations, 
that too either in the inviscid limit (Lamb \cite{lamb1993hydrodynamics}) 
or the asymptotic limit of vanishing viscosity (Prosperetti \cite{prosperetti1980free,prosperetti1981motion}). 
For our purposes, we use the configuration of the viscosity-damped capillary oscillations 
of a planar interface, as implemented by Popinet \& Zaleski \cite{popinet1999front}.  


\begin{figure}[h!]
    \centering
    \includegraphics[width = 0.9\textwidth]{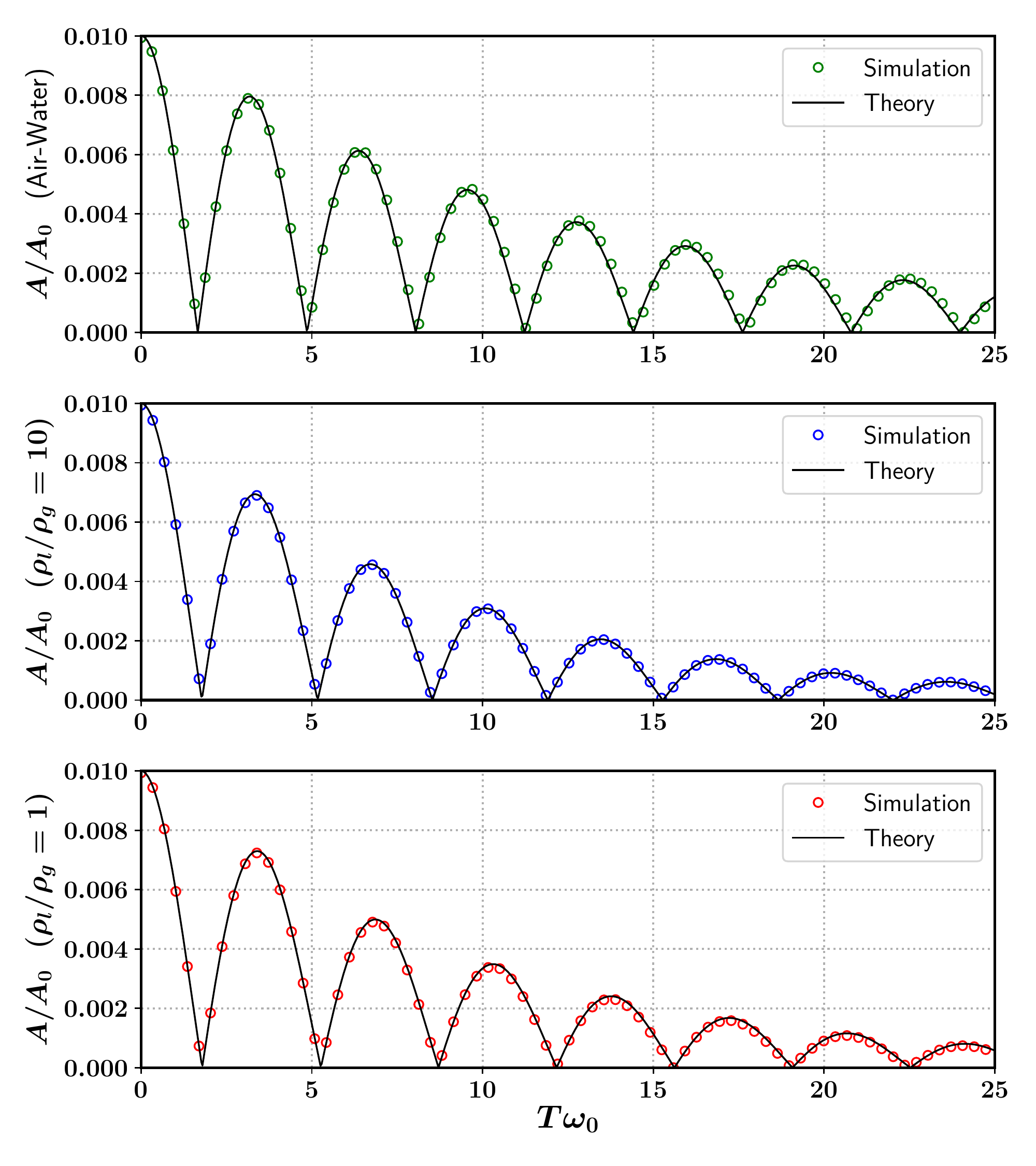}
	\caption{\textbf{STD} : Time evolution of the amplitude of the planar interface undergoing damped capillary oscillations, comparing the solution obtained by the non-consistent method with the closed-from Prosperetti solution. A relatively good agreement with theory is observed for all except the largest density-ratio tested. }
    \label{capwave_nonmc}
\end{figure}

\begin{figure}[h!]
    \centering
    \includegraphics[width = 0.9\textwidth]{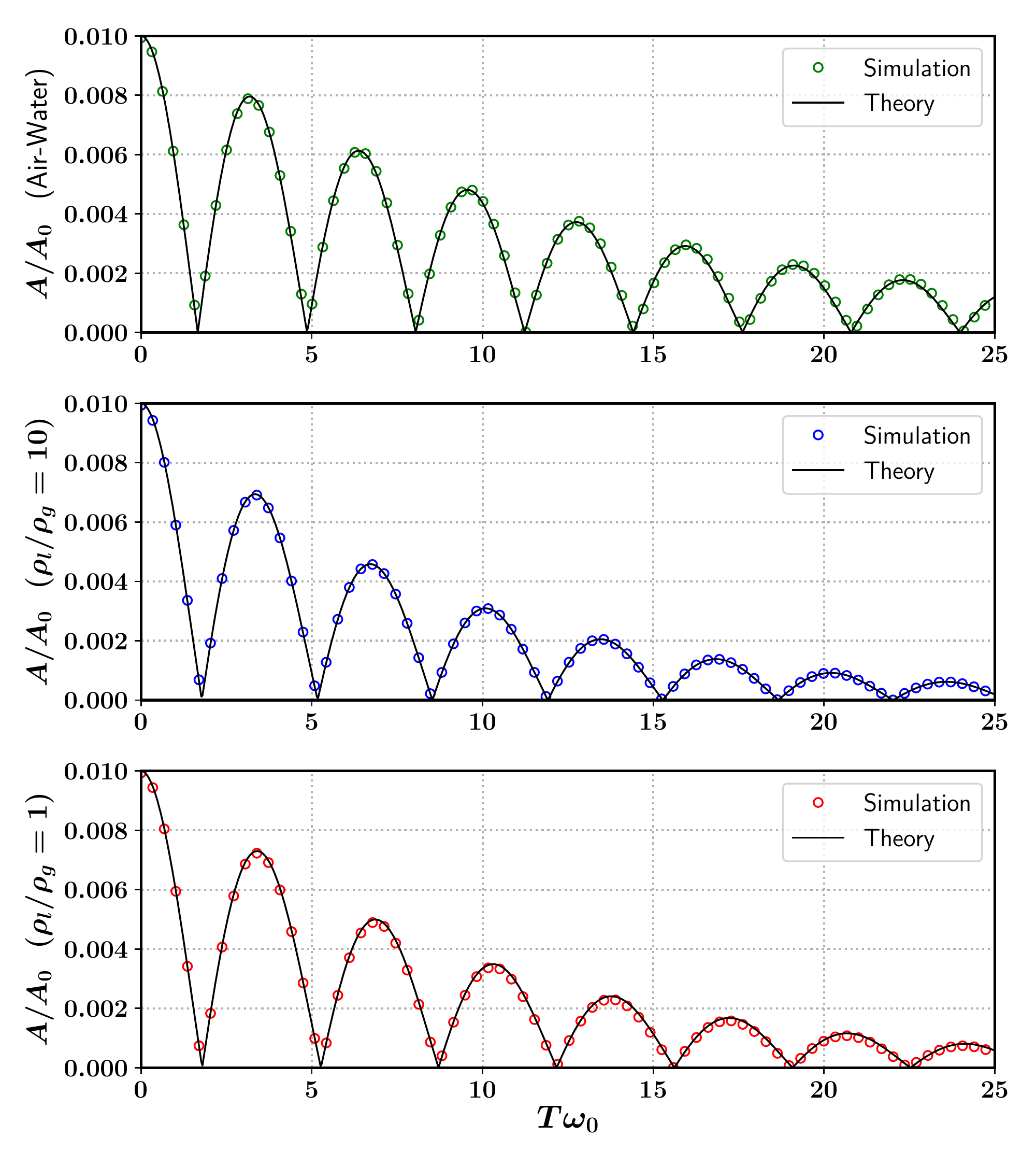}
	\caption{\textbf{MSUB} : Time evolution of the amplitude of the planar interface undergoing damped capillary oscillations, comparing the solution obtained by the consistent and conservative method with the closed-from Prosperetti solution. A marginally better agreement with theory is observed for all the density-ratios tested, compared to \textbf{STD} especially for the largest density-ratio. }
    \label{capwave_sagar}
\end{figure}

We consider a rectangular domain of dimensions $L \times 3L$, where $L$ corresponds 
to the wavelength of our initial perturbation. 
The densities of the heavier and lighter phases are $\rho_l$ and $\rho_g$ respectively, 
likewise for the viscosities $\mu_l$ and $\mu_g$, and $\sigma$ 
being the surface tension coefficient (fig. \ref{capwave_conf}). 
An intial perturbation amplitude of $L/100$ is used, coupled with a 
numerical resolution of $L/\Delta x= 32$ . 
Symmetry conditions are applied on the top and bottom sides, 
with periodic conditions along the horizontal direction. 
The problem is characterized by the adimensional parameters given as  

\begin{align}
	T_0 = T \omega_0 \quad , \quad \textrm{La} = \frac{\rho_l \sigma L}{\mu_l^2} \,, 
\end{align}

where $\textrm{La}$ is the Laplace number based on the heavier fluid, 
and $\omega_0$ is defined using the dispersion relation \cite{popinet2009accurate} given as  

\begin{align}
	\omega_0^2 =  \frac{\sigma k^3}{2 \rho_l} \quad, \qquad \text{where} \quad k = \frac{2\pi}{L} \,. 
\end{align}

The above relation is obtained using linear stability analysis at the 
inviscid limit \cite{lamb1993hydrodynamics}. 
In order to evaluate the influence of density contrast on the performance of our method, 
we use 3 different numerical setups keeping the same Laplace number ($\textrm{La} = 3000$)  

\begin{itemize}
	\item $\rho_l/\rho_g = 1$ , $\mu_l/\mu_g = 1$  (Popinet \cite{popinet2009accurate}). 
	\item $\rho_l/\rho_g = 10$ , $\mu_l/\mu_g = 1$  . 
	\item $\rho_l/\rho_g = 1000.0/1.2$ , $\mu_l/\mu_g = 1.003\cdot 10^{-3}/1.8\cdot 10^{-5}$ (Air-Water) . 
\end{itemize}

The final setup corresponds to that of an air-water interface, 
which is also the most stringent due to the large density and viscosity contrasts.  

\begin{figure}[h!]
    \centering
    \includegraphics[width = 0.9\textwidth]{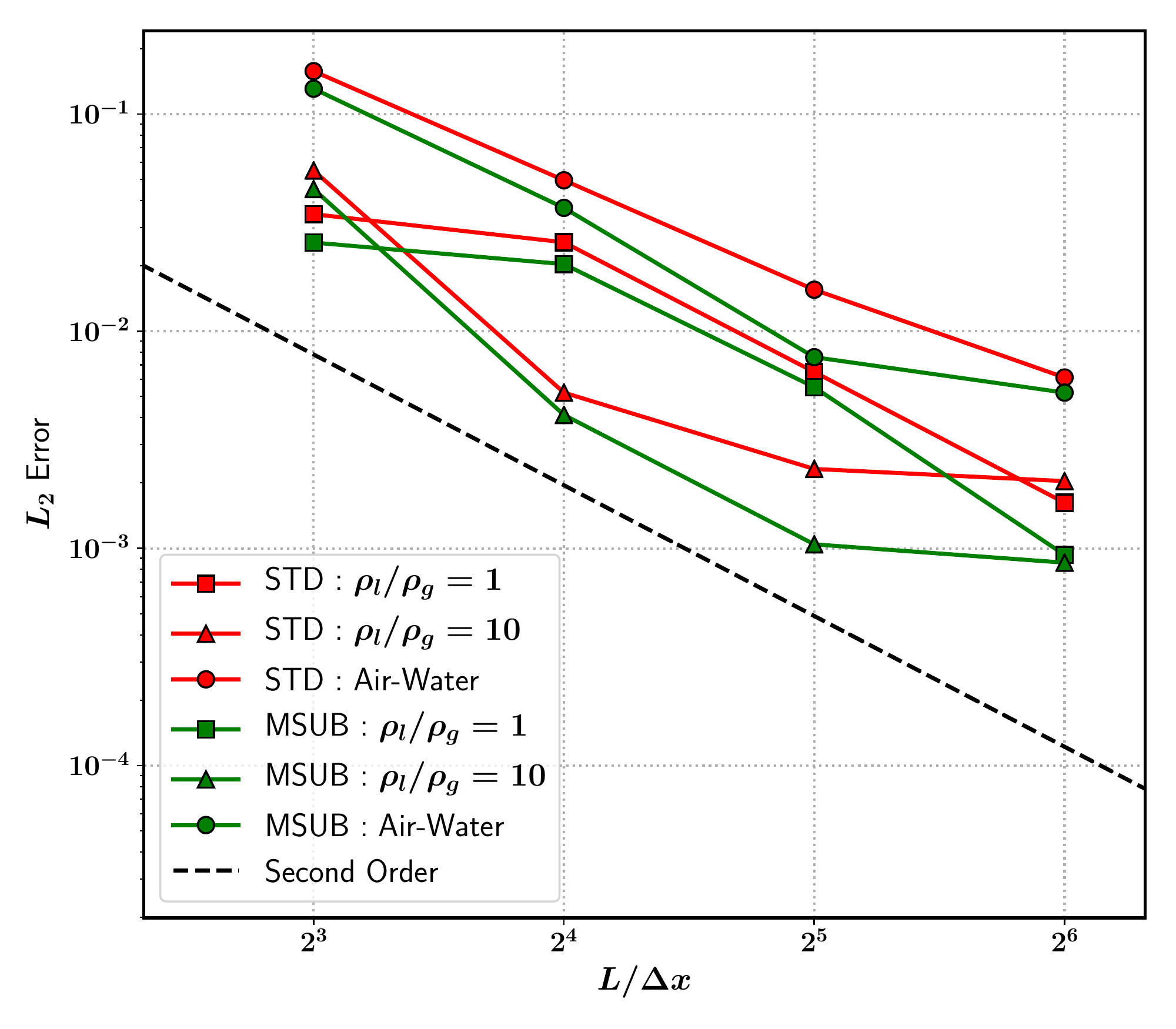}
	\caption{Comparison of spatial convergence for all three setups, between the non-consistent and the present method. Both methods seem to demonstrate approximately second-order convergence. The consistent and conservative method (\textbf{MSUB}) delivers slightly lower errors in all the setups tested, although there is some saturation in the convergence rate at higher resolutions for the stringent air-water setup. }
    \label{conv_all}
\end{figure}

The analytical solution for this configuration corresponds to 
closed-form expressions of the planar interface shape evolution 
given by Prosperetti \cite{prosperetti1981motion,prosperetti1980free}, 
which takes into account the finite time-scales at which the vorticity 
(generated due to interface oscillations) diffuses into the bulk medium. 
These expressions are numerically solved using using a fourth-order Runge-Kutta time integrator.
The amplitude is normalized by the initial value ($A_0$) and the time rescaled by $T_0$. 
As we can in figures \ref{capwave_nonmc} (\textbf{STD}) and \ref{capwave_sagar} (\textbf{MSUB}), 
both numerical methods provide relatively good agreement with the Prosperetti solution (black curves).
There is hardly any appreciable qualitative difference between the 
results obtained via the different methods \textbf{STD} and \textbf{MSUB}, 
although the present method (\textbf{MSUB}) seems to perforn marginally better 
for the  most stringent case (air-water configuration). 
Next, we evaluate the accuracy of our numerical results to the 
Prosperetti solution using an integral (in time) error norm, 
the same as defined in Popinet \cite{popinet2009accurate}, given as        

\begin{align}
	L_2 = \frac{1}{L} \sqrt{\frac{\omega_0}{25} \int_{t=0}^{T} \left(h - h_\textrm{exact}\right)^2} \,,
\end{align}

where $h$ is the maximum inteface height obtained using our numerical simulations, 
and $h_{exact}$ being the maximum height obtained via time integration of the Prosperetti solution. 
In Fig. \ref{conv_all} we show the rate of spatial convergence of the $L_2$ 
error norms for different density contrasts, simultaneously comparing the behavior of 
the different methods \textbf{STD} and \textbf{MSUB}. 
We maintain $\textrm{La} = 3000$ for all density contrasts, spatial resolutions and methods tested. 
We observe roughly second-order spatial convergence when it comes to 
equal densities across the interface, with \textbf{STD} and \textbf{MSUB} 
displaying nearly identical behavior, although \textbf{MSUB} performs 
slightly better with lower ($L_2$) errors for all resolutions. 
When it comes to $\rho_l / \rho_g = 10$ , we observe a saturation in 
the initial second-order convergence rate irrespective of whichever method is used, 
however \textbf{MSUB} again has lower errors. 
Finally, both methods demonstrate roughly second-order convergence when 
it comes to the stringent air-water configuration, with \textbf{MSUB} performing better 
with lower errors at the moderate resolutions. 
Not surprisingly, the largest errors arise for the air-water 
configuration errors irrespective of the method.

\section{Falling Raindrop}
\label{sec:raindrop}
We turn our attention towards the issues plaguing standard non-consistent numerical methods 
when dealing with realistic flows involving large density contrasts 
(e.g air and water systems), primarily in the form of droplets and bubbles.
A flow configuration that combines the difficulties of large 
density contrasts in conjunction with complexities involved 
due to the coupling of inertia, capillary, viscous stresses 
is that of a water droplet falling in air subject to gravity.

\subsection{Problem Setup}

The problem is characterised by a combination of Reynolds, 
Weber and Bond numbers, the definitions of which are as follows : 

\begin{align}
	\textrm{We}=\frac{\rho_{g} U^2 D}{\sigma} \quad,\quad \textrm{Re}= \frac{\rho_{g} U D}{\mu_{g}} \quad,\quad \textrm{Bo}=\frac{\left(\rho_{l}-\rho_{g}\right) g D^2 }{\sigma} \,.
\end{align}

\begin{figure}[h!]
\begin{center}
\includegraphics[width=0.6\textwidth]{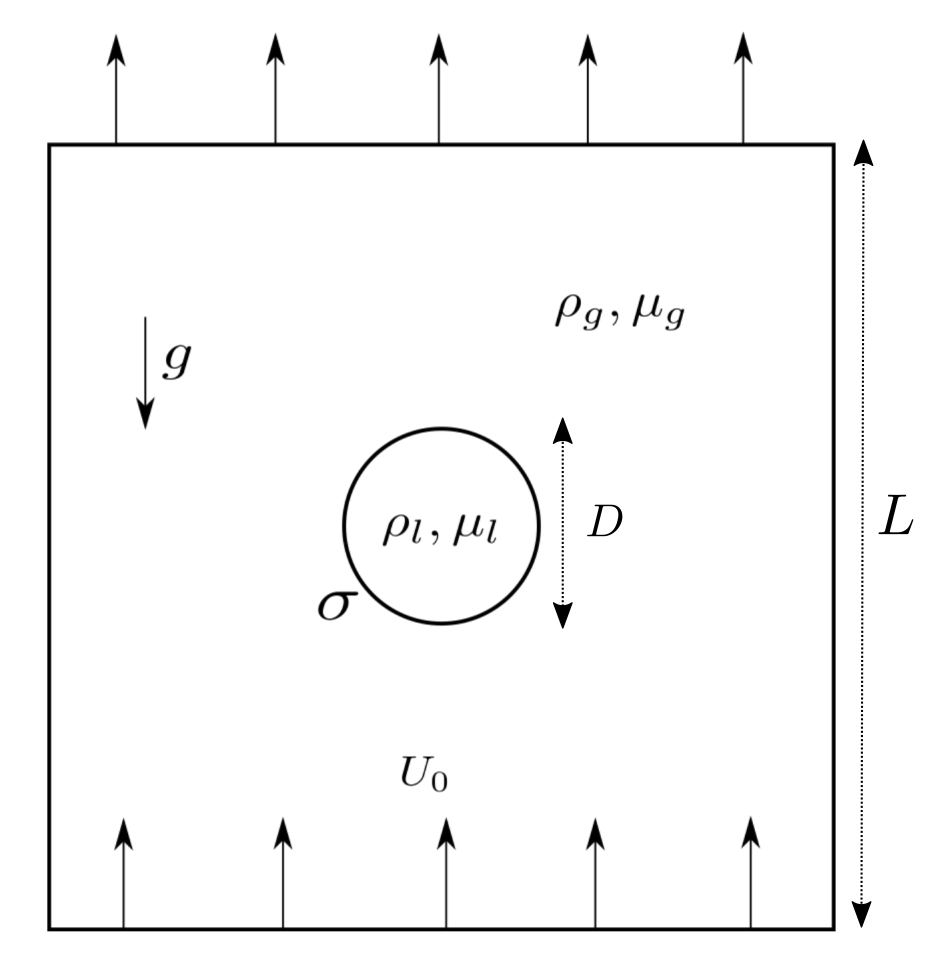}
\end{center}
\caption{A 2D schematic of the falling raindrop computational setup. 
	A droplet of diameter $D$ is placed at the center of a cubic domain 
	of side $L$ and $L/D = 4$. The liquid properties ($\rho_l$ , $\mu_l$) 
	correspond to that of water, and the gas properties ($\rho_g$,$\mu_g$) 
	correspond to that of air. We apply a uniform inflow velocity condition 
	with $U_0$ and an outflow velocity condition at the top which 
	corresponds to zero normal gradient.
	Boundary conditions on the side walls correspond 
	to those of impenetrable free slip (zero shear stress).}
\label{setup}
\end{figure}

The density and viscosity ratios are corresponding to that of 
air-water systems at 20 degree Celcius.
An equivalent characterization could be done using the 
Morton number $\left( \textrm{Mo} = \frac{g\mu_g^4(\rho_l-\rho_g)}{\rho_g^2 \sigma^3} \right)$ 
instead of the Bond number.
The subscripts $l$ and $g$ represent liquid and gas phases respectively. 
In our particular numerical setup, $\textrm{We}\simeq 3.2 $, 
$\textrm{Re} \simeq 1455 $ and $\textrm{Bo} \simeq 1.2 $,
thus corresponding to that of a $3mm$ diameter raindrop (a relatively large one) 
falling in the air at an approximate terminal velocity of  
$8$ m/s (interpolated from empirical data, refer to Gunn and Kinzer \cite{gunn1949}). 
This choice of length scale of the drop is motivated by the paradigmatic value
of a near-spherical raindrop simulation, and by the fact that the corresponding Weber
number ($\sim 3$) is the same as in a similar air-water setup corresponding to  
suddenly-accelerated-droplet (or ``secondary atomization" ) simulations in the studies \cite{xiao2012,xiao2014large}.
For such a low Weber number, the capillary forces should keep the interface shape more or less intact.
The parameters in the problem setup are given in Table \ref{raindropprop}, 
and a schematic diagram given in Fig. \ref{setup}. 
The droplet is initially placed at the center of a cubic domain (3D), 
where the length of the side is 4 times the diameter of the drop. 

\begin{table*}[h!]
\begin{center}
\begin{tabular}{ccccccc}
\hline\hline
$\rho_{g}$ & $\rho_{l}$ & $\mu_{g}$ 
& $\mu_{l}$ & $\sigma$ & $D$ & $g$\\
$\left(kg/m^3\right)$ & $\left(kg/m^3\right)$ & $\left(Pa \, s\right)$ 
& $\left(Pa \,s \right)$ & $\left(N/m\right)$ & $(m)$ & $(m /s^{2})$ \\
\hline
1.2 & $0.9982 \times 10^3$ & $1.98 \times 10^{-5}$ & 
$8.9 \times 10^{-4}$ & $0.0728$ & $3 \times 10^{-3}$ & $9.81$\\
\hline\hline
\end{tabular}
\caption{Parameter values used in the simulation 
	of a falling water droplet in air. \label{raindropprop}}
\end{center}
\end{table*}

In order to properly reproduce and analyse the dynamics 
of a relatively large drop (high Reynolds flow) such as in our case, 
the numerical method has to accurately resolve the thin boundary layers,
the interaction of such layers with the capillary forces, and finally 
the non-linear feedback of the complex 3D vortical structures in the wake.
In our experience, even for simulations with 64 points across the droplet diameter, 
the resulting boundary layer has only by 3-4 cells across it.
Arguably, the most natural type of computational setup would involve using 
a large domain filled with air at rest, with zero inflow velocity and to
let the droplet fall from the top of the domain.
Such an undertaking was attempted by Dodd and Ferrante \cite{dodd2014}, 
in which they managed to delineate the different regimes concerning the 
behavior of the wake behind the droplet, although at relatively 
low Reynolds numbers corresponding to smaller drops 
(the maximum Reynolds tested was $\simeq 500$, whereas in our case it is $\simeq 1500$).    

We use a significantly smaller domain ($L/D = 4$ where $L$ is the domain size), 
with a constant value of inflow velocity (close to $8$ m/s), 
implying that the drop will exit the domain after some time. 
This setup proves to be quite convenient and computationally efficient 
for relatively short-time investigations.
Therefore, our objective behind the demonstration of this particular 
case is \textit{not} to develop a high fidelity model of a raindrop, 
but instead carry out a stringent evaluation of the robustness of our 
present method compared to the standard non-consistent version. 


\subsection{Present Method vs. Non-consistent method}

\begin{figure}[h!]
\begin{center}
\includegraphics[width = 1.0\textwidth]{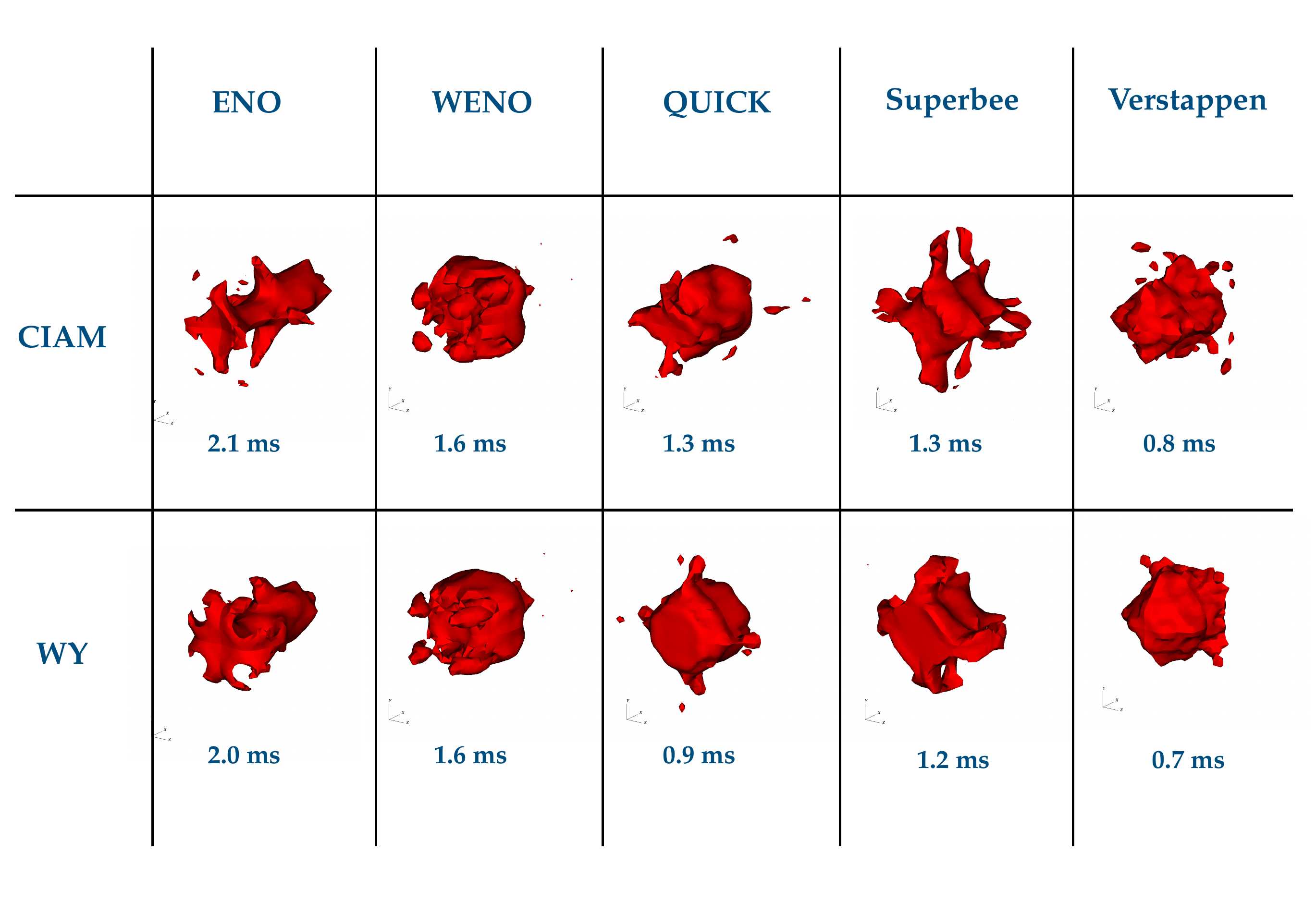}
\end{center}
\vspace*{-0.5cm}
\caption{A catalogue of the different manifestations of
``artificial'' atomization encountered while using the standard
non-consistent method (\textbf{STD}).
The red contours indicate the isosurface of the volume fraction
field corresponding to a value of $0.5$, which is a good proxy for the exact interfacial position.
The resolution of the droplet in all the cases is $D/ \Delta x = 16$.
The symbol ``WY'' in the legend corresponds to those run using
the Weymouth-Yue advection scheme, and ``CIAM'' corresponds to the a Lagrangian explicit 
flux advection scheme, as detailed in \cite{paris,caf2020,gueyffier}.
The implementation of the non-linear flux limiters i.e WENO, ENO, QUICK, Superbee, Verstappen
are identical to that of well established methods in the context of hyperbolic conservation laws, 
the reader can refer to \cite{flim_1,flim_2,caf2020} for more details.
The time stamps are indicative of the moments at which the interface
appears most deformed, and do not necessarily refer to moments at which the solver crashes.
We can clearly observe that the un-physical fragmentation of the raindrop is symptomatic
of the non-consistent method (\textbf{STD}), systematically across 
all combinations of flux limiters and advection schemes.
}
\label{crashes}
\end{figure}

\begin{figure}[h!]
\includegraphics[width=0.8\textwidth]{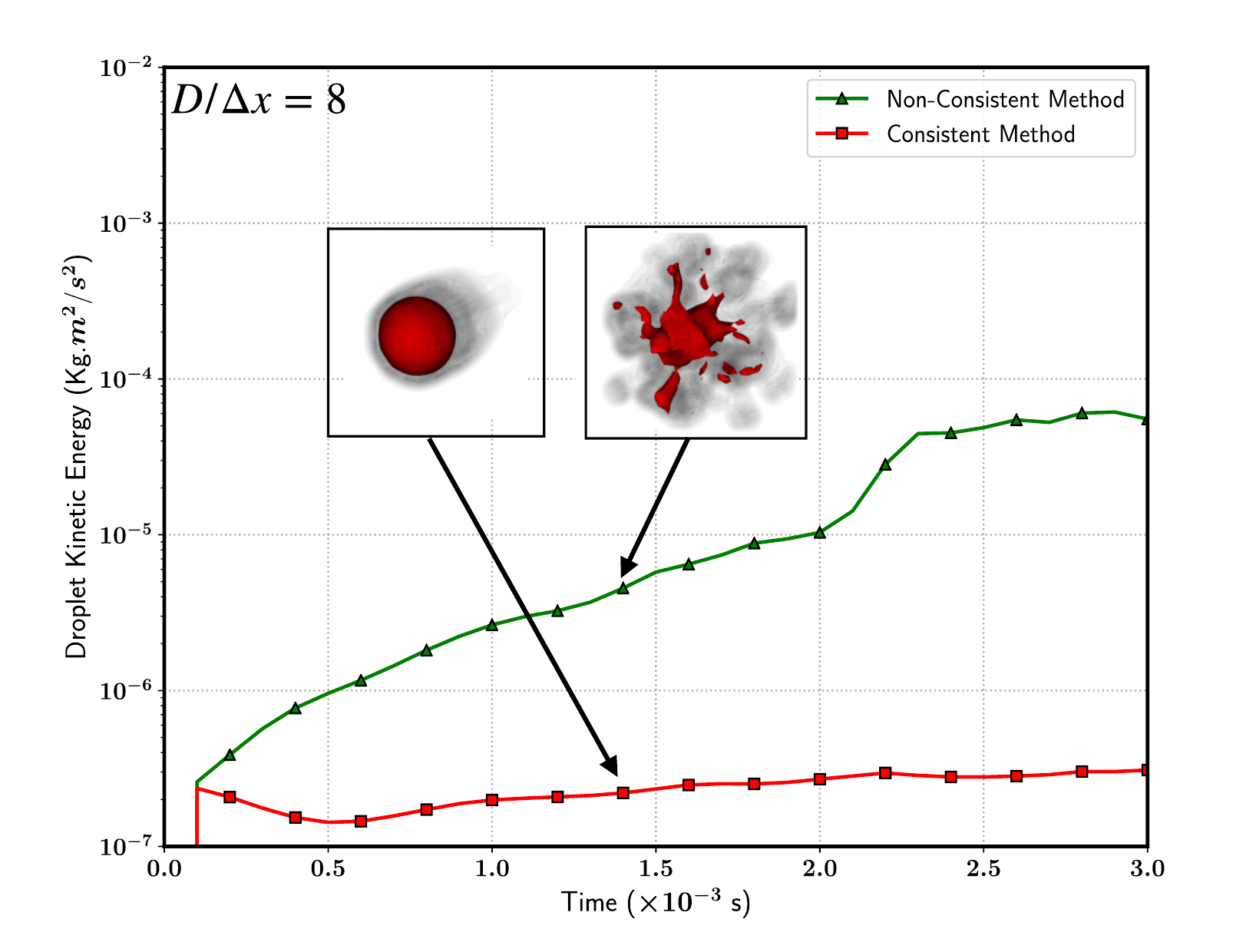}
\centering
\caption{
Comparison of the temporal evolution of droplet kinetic energy
between the non-consistent (\textbf{STD}) method and the present (\textbf{MSUB}) method.
The kinetic energy is computed according to \eqref{drop_ke},
and the droplet resolutions are $D/\Delta x = 8 $ for both sets of simulations.
In the case of the non-consistent method, the kinetic energy of the drop 
climbs climbs rapidly over several orders of magnitude before 
the drop finally breaks up into many smaller fragments, 
In contrast, kinetic energy of the drop evolves 
in a smooth manner using our present method. 
}
\label{explode_compare}
\end{figure}

Numerical simulations using a fixed inflow velocity with $U_0 = 8 $ m/s 
were carried out for very short times (a few milliseconds) at moderate resolution 
corresponding to $D/\Delta x = \{8,16,32,64\}$ , 
where $D$ is the diameter and $\Delta x$ the grid size. 
Simulations carried out using the standard non-consistent version (\textbf{STD}) 
result in catastrophic deformations of the droplet as catalogued in Fig. \ref{crashes}, 
which we describe as ``fictitious" or ``artificial" atomization. 
They display marked peaks or spikes in kinetic energy as a function of time, 
associated with massively deformed interface shapes (see figure \ref{explode_compare}). 

A simplified model based on inviscid steady-state flow 
presented in our previous study \cite{caf2020}
explains the origins of this numerical instability.
In essence, simulations carried out with the standard method 
(for moderate resolutions) lead to mixing of the gas
velocity with the liquid density at mixed cells near 
the hyperpolic stagnation point at the droplet front.
This results in extremely large (unphysical) pressure gradients 
which can only be balanced by the surface tension force 
by invoking sufficiently large curvatures. 
These pressure ``spikes'' (see \cite{caf2020}) across the interface 
eventually lead to its rapid destabilization and concomitant breakup.  

\begin{figure}[h!]
\includegraphics[width=1.0\textwidth]{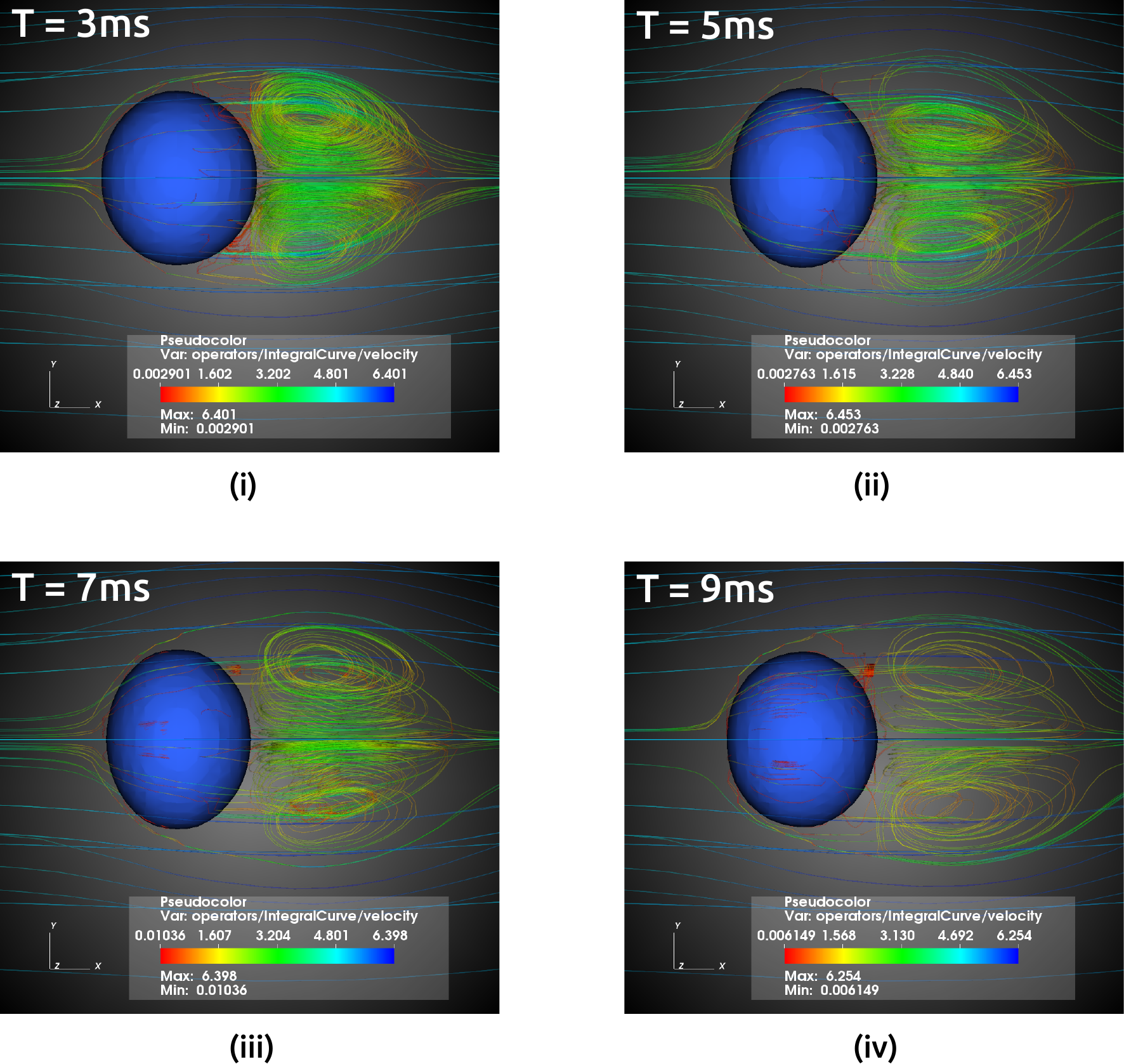}
\centering
	\caption{Numerical simulations using our present method (\textbf{MSUB}) 
	of a $3 mm$ raindrop falling in air under gravitational acceleration.
        The flow is along the positive X direction with a constant inflow velocity of 8m/s, 
	gravity is along the opposite direction, and with ``QUICK'' as the flux limiter.
        The blue contour indicates the isosurface of the volume fraction
        field corresponding to a value of $0.5$, and the instantaneous streamlines
	are colored according to the velocity magnitude. 
        The droplet has a relatively poor numerical resolution of $D/\Delta x = 16$ . 
        Within the short time scales under consideration, the droplet shape undergoes oscillations 
	(corresponding to the $2^{\textrm{nd}}$ Lamb mode \cite{lamb1993hydrodynamics})
	along the flow direction, with the structure of the wake becoming more chaotic with time. 
	This demonstrates that our present method (\textbf{MSUB}) can easily suppress the numerical instabilties
	and massive interfacial deformations that are rampant in the case of the standard (\textbf{STD}) method.
        }
        \label{msub_series}
\end{figure}

The most common and brute force approach that one can apply 
in order to suppress or circumvent such numerical instabilities is 
by using a combination of extremely refined meshes coupled with large domains \cite{dodd2014}.
In our study, we have observed that even a resolution of $D/\Delta x = 64$ is not 
sufficient to suppress the instabilities in case of the standard method, although
some combinations of advection schemes (CIAM, Weymouth-Yue) and 
flux limiters (WENO,ENO,Superbee etc) seem to be more stable than others. 
We observe that the use of our present method enables us to stabilize the 
aforementioned numerical instabilities (compare Fig. \ref{msub_series} and Fig. \ref{crashes}), 
that too in a systematic manner that spans a wide range of flux 
limiters (WENO, ENO, Superbee, QUICK, Verstappen) and CFL numbers. 
The kinetic energy of the droplet (using \textbf{MSUB}) evolves in a relatively smooth manner, 
without the presence of sudden spikes and falls which are emblematic of 
the non-consistent method (refer to Fig. \ref{explode_compare}). 
Such abrupt changes in kinetic energy of the droplet have been 
found to be associated with instants when the droplet undergoes 
``artificial'' atomization or breakup, henceforth resulting in 
catastrophic loss of stability for the non-consistent (\textbf{STD}) method. 

\subsubsection{Convergence Study}

For the next set of simulations, we use a fixed inflow velocity setup but with smaller initial velocity.
We systematically vary the resolution from  $D/\Delta x = 8, 16, 32 $ and $64$. 
Despite using our present method, simulations at $D/\Delta x = 8$ are sometimes unstable, 
so we use a workaround and use a lower fixed inflow velocity of $U_0=5$ m/s. 
This simplification acts as a milder initial condition and enables us 
to observe the first phase of the (physical) acceleration towards the final statistical steady state. 
The simulations are carried out for 5 ms, in order to minimize computational cost as well as to 
avert situations in which droplet approaches too close to the domain boundaries. 
There are several time scales in this problem, the important ones are

\begin{itemize}
\item{$t_a=D/U_0 \approx 6$ ms : }The time scale for the air to flow around the droplet.
\item{ $t_w = L/[2(U_t - U_0)] = 3$ ms : }The time taken by the drop to approach the domain boundary, assuming terminal velocity. 
\item{$t_c \simeq 15.1$ms : }The time scale of capillary oscillations \cite{rayleigh1879a,lamb1993hydrodynamics} of the droplet shape.
\item{$t_i = (\rho_l/\rho_g)\cdot( D/U_t) = 215$ ms : }The time scale to reach statistically stationary state (terminal velocity), 
This estimate is obtained using a simple 1D model of the droplet dynamics (refer to \cite{caf2020}) using a square-velocity drag law. 
\item{$t_\mu = \textrm{Re}\cdot (D/U_t) = 400 $ms : } The time scale for the gas inertia to entrain vortical motion inside the drop (liquid side) due to the action of shear at the interface. 
\end{itemize}

\begin{figure}[!h]
\begin{center}
\includegraphics[width = 1.0\textwidth]{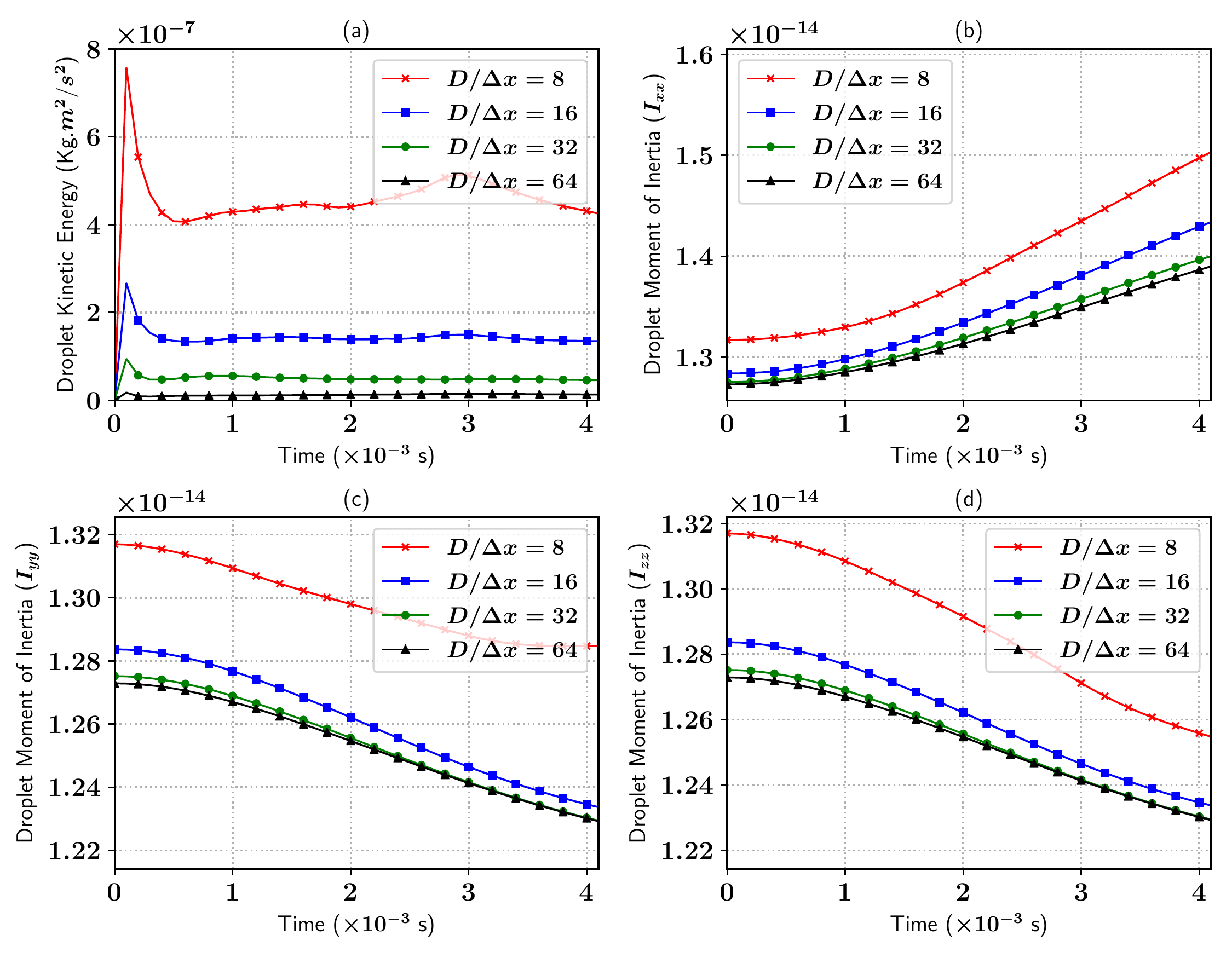}
\end{center}
\vspace*{-0.5cm}
\caption{Temporal evolution of quantities of interest to evaluate the 
	performance of our consistent scheme based on the 
	sub-grid (\textbf{MSUB}) method, for different spatial resolutions. 
	(a) Kinetic energy relative to the droplet center-of-mass as defined in \eqref{drop_ke}. 
	(b) Moment of inertia of the droplet along the flow (X) direction. 
	(c) and (d) Moment of inertia of the droplet along the directions 
	perpendicular to flow (Y,Z), evolution of $I_{yy}$ 
	seems to be more or less identical to $I_{zz}$.}
\label{multi_jcp}
\end{figure}

\begin{figure}[!h] 
\begin{center}
\includegraphics[width = 1.0\textwidth]{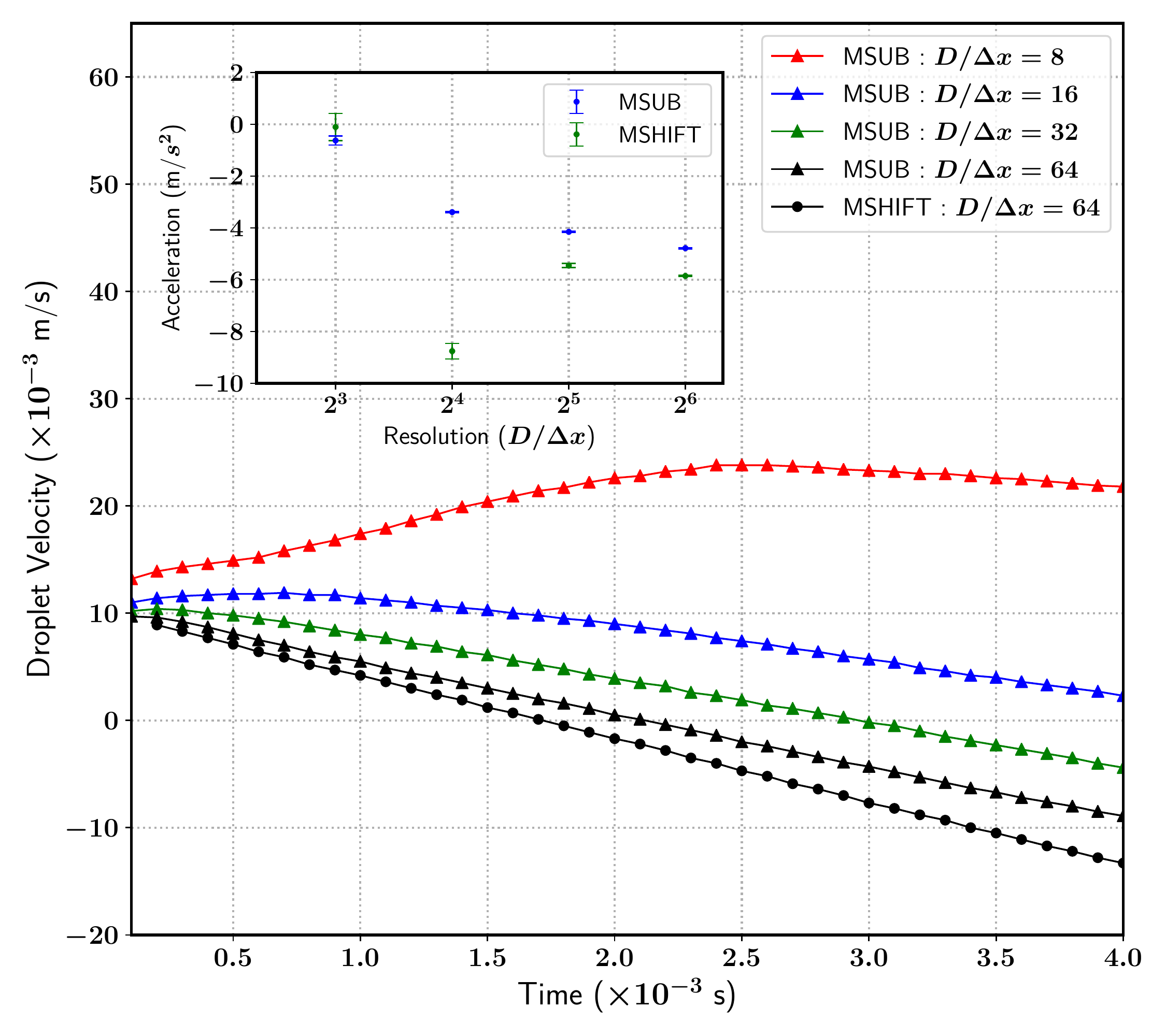}
\end{center}
\vspace*{-0.5cm}
	\caption{Comparison of droplet (center of mass) velocity as a function of time, for different droplet resolutions.
	The simulations were carried out with a consistent scheme based on the 
	present (\textbf{MSUB}) method, using the WY advection scheme with the QUICK limiter. 
	For comparison with our previous study \cite{caf2020}, we have added the results obtained 
	with the \textit{shifted fractions} method (\textbf{MSHIFT}, WY and QUICK), 
	corresponding to $D/\Delta x = 64$. 
	Inset : Droplet acceleration estimates as a function of its resolution,  
	computed using the best linear fit over the interval $[1\textrm{ms},4\textrm{ms}]$. 
	The error bars give the asymptotic standard error of the least-squared fits.} 
\label{drop_vel_jcp}
\end{figure}

We show the results of our convergence study in figures \ref{multi_jcp} and \ref{drop_vel_jcp}.
The quantities of interest while examining the robustness of the 
present method are the temporal evolution of the droplet kinetic energy 
(Fig. \ref{multi_jcp}. (a) ) , and the moments of inertia of the droplet 
along the three directions (Figures \ref{multi_jcp}. (b),(c),(d)) . 
The inflow is along the X direction with gravity opposite to it. 
The moment of inertia is used as a descriptor of the ``average'' shape of the drop, 
with their definitions along the different axes $I_m$ given by  

\begin{align}
	I_m = \int_{\mathcal{D}} \chi x_m^2 d \boldsymbol{x} \quad , \quad \text{where}, \quad 1 \le m \le 3 \,,
\end{align}

where $\mathcal{D}$ is the computational domain and $x_m$ is the 
distance of the interface relative to the center of mass of the droplet.   
The droplet kinetic energy is defined relative to the droplet center-of-mass, given by 
 
\begin{align}
	E_k = \langle \rho_l \cdot \chi(x,t) || \boldsymbol{u}(x,t) - \boldsymbol{u}_{CM}||^2 \rangle \,,
	\label{drop_ke}
\end{align}

where $ \langle . \rangle$ is the spatial averaging operator over the  
entire domain and $\boldsymbol{u}_{CM}$ is the droplet center of mass.
We observe a systematic drop in the amount of the droplet 
kinetic energy as we increase resolution, with the most probable explanation 
being that of the suppression of spurious interfacial 
jitter which is rampant at lower resolutions. 

\begin{figure}[h!] 
\begin{center}
\includegraphics[width = 0.7\textwidth]{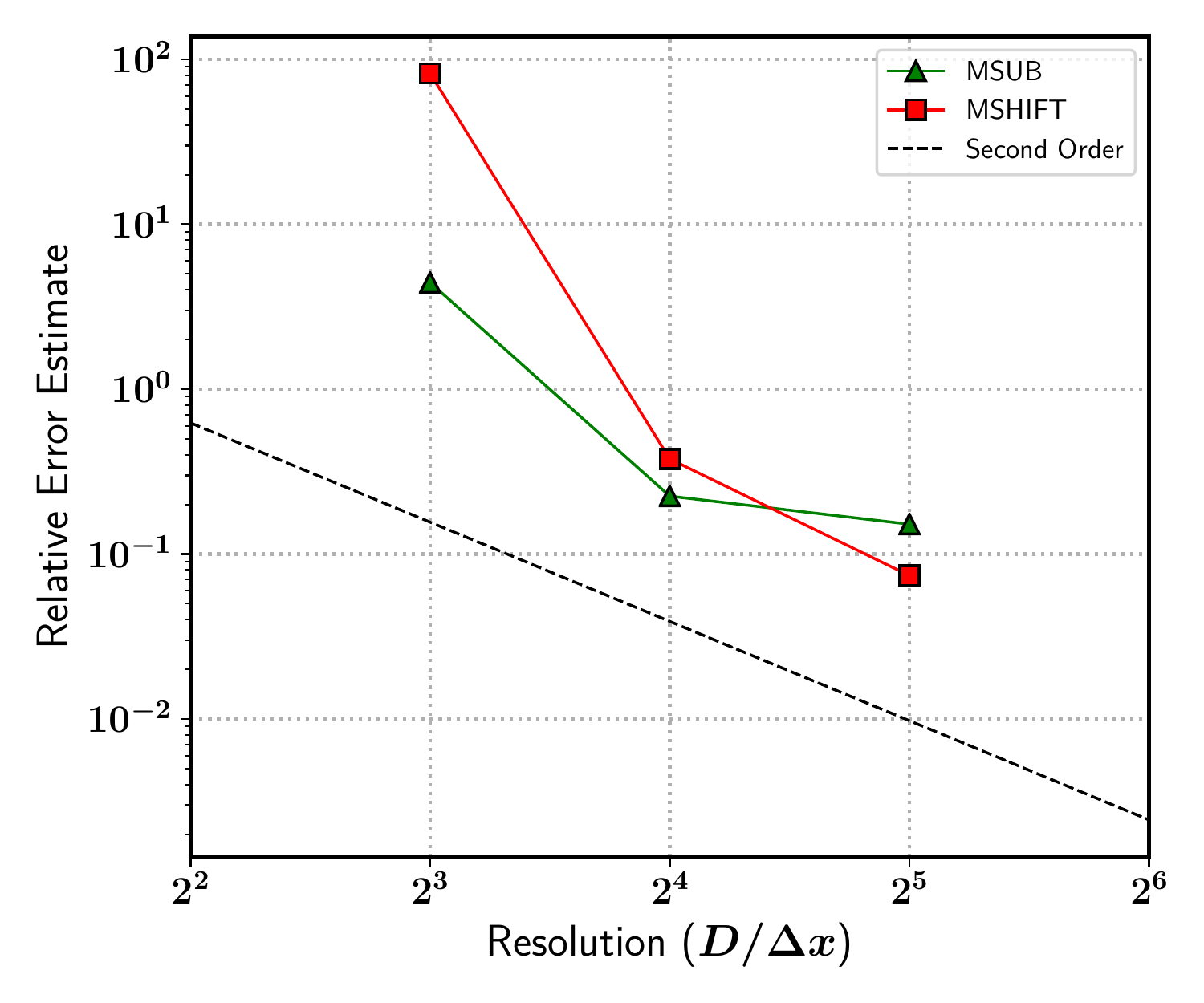}
\end{center}
\vspace*{-0.5cm}
	\caption{Variation in the error estimates of droplet acceleration as a function of 
	droplet resolution. The \textbf{MSHIFT} method displays a better convergence rate,
	but that is due to the wildly oscillating behaviour when it comes to the acceleration estimate
	as a function of resolution. The \textbf{MSUB} method on the other hand, displays a monotonic 
	decrease in the acceleration estimate with resolution, although with a saturating convergence rate.
	The second order convergence rate is plotted as a reference.} 
\label{drop_acc}
\end{figure}

There is also a component of the kinetic energy of the droplet 
associated with the internal coherent vortical structures generated due to 
the interaction with aerodynamic shear at the interface, 
evidenced by the non-zero value of the kinetic 
energy even for the most highly resolved droplets. 
Finally, the moments of inertia of the droplet evolve in a smooth manner,
and seem to converge as we go from $D/\Delta x=32$ to $D/\Delta x =64$,
for each of the 3 directions.

In fig. \ref{drop_vel_jcp}, we plot the the velocity 
of the center of mass of the droplet as a function of time, 
and its behavior for increasing droplet resolutions. 
The temporal variation in the droplet velocity is fitted to a 
linear polynomial in order to evaluate the droplet acceleration 
by means of a standard least-squares approach. 
The acceleration corresponding to the finest droplet resolution 
is $\frac{\textrm{d}U}{\textrm{d}t} \simeq 4.5 \pm 0.4$ $m/s^2$
, where the uncertainty is estimated from the difference between   
$D/\Delta x = 2^5$ and the $D/\Delta x = 2^6$ .
The acceleration estimate obtained from our previous study \cite{caf2020} 
which uses the \textit{shifted fractions} (\textbf{MSHIFT}) method with an identical setup, 
is $\frac{\textrm{d}U}{\textrm{d}t} \simeq 5.8 \pm 0.1$ $m/s^2$.
The discrepancy between the acceleration estimates are clearly visible in the inset of fig. \ref{drop_vel_jcp},
where the \textit{shifted fractions} method displays some oscillatory behaviour with increasing resolution. 
The present method (\textbf{MSUB}) on the other hand displays a monotonic trend towards a converged estimate, 
and also has lower fitting errors compared to \textbf{MSHIFT} (inset fig. \ref{drop_vel_jcp}), indicating that
the \textbf{MSHIFT} method is ``noisier'' than our present method.

Fig. \ref{drop_acc} shows the convergence rate of the methods.
A relative error estimate at resolution level $n$ (i.e. $D/\Delta x = 2^n$) is defined as the 
difference between the acceleration estimates (inset fig. \ref{drop_vel_jcp}) at the $n$ and $n+1$ levels,
normalized by the value at level $n$.  
The underlying cause of the difference between \textbf{MSUB} and \textbf{MSHIFT} 
methods is most likely due to the differing levels of dissipation between the methods. 
The \textit{shifted fractions} method is more dissipative, which might lead to faster (and greater) momentum diffusion.
This might explain the differences we have observed in the (not shown) temporal development 
of the boundary layers around the raindrop, between the \textbf{MSUB} and \textbf{MSHIFT} methods. 
A more detailed and quantitative anylysis is warranted to verify the above assertions, 
but it is beyong the scope of our present study.
To sum it up, the results obtained in the case of the falling droplet strongly suggest that 
our present numerical method (\textbf{MSUB}) can be used to get relatively good 
estimates of the underlying flow features of the drop, without observing 
any un-physical evolution due to the discretization errors at low to moderate resolutions.

\section{Conclusions \& Perspectives}
We have presented a numerical method designed to simulate complex liquid-gas 
interfacial flows involving significant contrasts in density. 
The conservative formulation of governing equations are solved 
with momenutum and density as the primary variables, on staggered Cartesian grids.
Rudman's \cite{rudman1998volume} strategy of mass advection on a doubly 
refined grid is adopted using geometric flux reconstructions so as to 
maintain consistent mass-momentum advection on the staggered grid.
The Weymouth and Yue algorithm \cite{wy} is extended to 
momentum transport, in order to ensure discrete consistency between 
the conservative direction-split (3D) transport of volume, mass and momentum.  

The performance of the method is assessed using standard benchmark cases 
such as the behavior of spurious currents in static and moving droplets, 
and the propagation of damped capillary waves. 
All of the test cases tests were conducted for density contrasts spanning 
3 orders of magnitude, including a realistic air-water configuration for capillary waves. 
Our method demonstrates increased accuracy compared to the standard non-consistent version. 
A second-order error convergence rate is observed for the static droplet and the capillary wave, 
with the rate reverting to first-order in the case of droplet advection.        

Finally, the robustness and stability of our present method was demonstrated 
using the case of a raindrop falling in air under gravitational acceleration. 
A key feature of the method is its ability to avoid un-physical interfacial deformations 
(artificial atomization of the raindrop), which are rampant in the case of the non-consistent method.   
The method delivers good estimates of the flow features especially at low resolution, 
as evidenced by the smooth (and convergent) evolution of the droplet kinetic energy and the moment of inertia. 
A short time investigation of the falling drop was able to reproduce the 
physically consistent initial acceleration phase (towards the final statistically stationary state) of the raindrop.

In terms of future perspectives, the present method can be further adapted to derive more accurate 
curvature estimates by using the volume fraction information available from the twice finer grid, 
and we are currently working towards finding the optimal strategy for the such curvature restrictions. 
The method is also going to be used to investigate several flow configurations of scientific interest
such as atomization of shear-layers and Kelvin-Helmholtz instabilties, secondary atomization of drops etc.   

\label{sec:conc}

\section*{Acknowledgements}
This work has been supported by the grant SU-17-R-PER-26-MULTIBRANCH
from Sorbonne Université and the ERC Advanced Grant TRUFLOW.
This work was granted access to the HPC resources of TGCC-
CURIE, TGCC-IRENE and CINES-Occigen under the allocations
A0032B07760 and 2020225418, made
by GENCI and PRACE respectively. 
We would like to thank G. Tryggvason, R. Scardovelli, Dr. Y.
Ling, Dr. W. Aniszewski, Dr. S. Dabiri, Dr. J. Lu and Dr. P. Yecko 
for their contribution to the development of the code ``PARIS Simulator''.
Finally, the simulation data are visualized by the software VisIt,
developed by the Lawrence Livermore National Laboratory.

\section*{Appendix A : \textit{Brief review of mass-momentum consistent methods}}
\label{sec:append_a}
A bird's eye view of the numerous features employed by the methods in existing literature, we refer the reader to tables \ref{table_vof} and \ref{table_ls}, which respectively provide systematic overviews of the VOF based and level set based approaches.    

\clearpage
  \begin{landscape}
      \begin{table}[p]
      	\renewcommand\arraystretch{2.5}
      	\raggedleft
	\caption{Summary of \textbf{Volume-of-Fluid} based mass-momentum consistent advection methods for incompressible flows of immiscible fluids.}
        \label{table_vof}
	  \resizebox{1.6\textwidth}{!}{%
	  \begin{tabular}[t]{ >{\bfseries\raggedright}m{0.15\linewidth} >{\raggedright}m{0.15\linewidth}  >{\raggedright}m{0.15\linewidth}  >{\raggedright}m{0.15\linewidth} >{\raggedright}m{0.15\linewidth} >{\raggedright}m{0.15\linewidth} >{\raggedright\arraybackslash}m{0.15\linewidth} }
      	  	\toprule
		  & Rudman \newline(IJNMF 1998) &  Bussmann et al. \newline (ASME 2002) & LeChenadec \& Pitsch \newline (JCP 2013) & Owkes \& Desjardins \newline (JCP 2017) & Patel \& Natarajan \newline (JCP 2017) & Arrufat et al. \newline (CAF 2020) \\
      	  	\midrule
		  
		  Basic Configuration & 2D Cartesian, \newline staggered & 3D hexahedral unstructured, \newline collocated & 3D Cartesian, \newline staggered & 3D Cartesian, \newline staggered & 3D polygonal unstructured, \newline staggered & 3D Cartesian \newline staggered \\ 

		  Interface Representation & VOF, Piecewise Linear & VOF, Piecewise Linear & VOF, Piecewise Linear & VOF, Piecewise Linear & VOF, none & VOF, Piecewise Linear \\

		  Flux Computation & split, algebraic Flux Corrected Transport, Eulerian  & unsplit, geometric, Eulerian & unsplit, geometric, semi-Lagrangian & unsplit, geometric, semi-Lagrangian & algebraic Cubic Upwind, Eulerian & split, geometric, Eulerian \\

		  Surface Tension & Continuum Surface-Force & not specified & Ghost-Fluid Method, well-balanced & Continuum Surface-Force, well-balanced & Continuum Surface-Force, well-balanced & Continuum Surface-Force, well-balanced \\
      	  	
		  Curvature Estimation & VOF convolution (smoothed) & not specified & VOF convolution (smoothed) & hybrid mesh-decoupled height functions & VOF convolution (smoothed) & hybrid height functions and curve fitting \\

                 Viscous Stresses & explicit, harmonic averaging & not specified & explicit, harmonic averaging & not specified & explicit, harmonic averaging & explicit, arithmetic averaging \\ 
		  
		  Pressure-Poisson Solver & preconditioned multigrid, Gauss-Seidel & not specified & not specified & not specified & preconditioned GMRES, Krylov subspace & preconditioned multigrid, Gauss-Seidel \\

      	  	\bottomrule
	  \end{tabular}}
      \end{table}
  \end{landscape}

\clearpage
  \begin{landscape}
      \begin{table}[p]
      	\renewcommand\arraystretch{2.8}
      	\raggedleft
	      \caption{Summary of \textbf{Level Set} and \textbf{CLSVOF} based mass-momentum consistent advection methods for incompressible flows of immiscible fluids}
	      \label{table_ls}
        \resizebox{1.6\textwidth}{!}{%
	  \begin{tabular}[t]{ >{\bfseries\raggedright}m{0.20\linewidth} >{\raggedright}m{0.16\linewidth}  >{\raggedright}m{0.16\linewidth} >{\raggedright}m{0.16\linewidth} >{\raggedright}m{0.16\linewidth} >{\raggedright\arraybackslash}m{0.16\linewidth} }
      	  	\toprule
		  & Raessi \& Pitsch \newline (CAF 2012)& Ghods \& Herrmann \newline (Physica Scripta 2013) & Vaudor et al. \newline (CAF 2017)
		  & Nangia et al. \newline (JCP 2019) & Zuzio et al. \newline (JCP 2020) \\
      	  	\midrule

		  Basic Configuration & 2D Cartesian, \newline staggered & 3D hexahedral unstructured, \newline collocated & 3D Cartesian, \newline staggered & 3D Cartesian \newline staggered & 3D Cartesian \newline staggered \\ 

		  Interface Representation & Level Set & Level Set & Coupled Level Set-VOF & Level Set & Coupled Level Set-VOF \\

		  Flux Computation & Level Set derived, semi-Lagrangian & Level Set derived , Eulerian & split, geometric, Eulerian & algebraic Cubic Upwind, Eulerian & split, geometric, Eulerian \\

		  Surface Tension & Ghost-Fluid Method & Continuum Surface-Force & Ghost-Fluid Method, well-balanced & Continuum Surface-Force, well-balanced & Ghost-Fluid Method \\
      	  	
		  Curvature Estimation & Level Set based & Level Set based & Level Set based & Level Set based & Level Set based \\

                 Viscous Stresses & implicit, harmonic averaging & explicit, arithmetic averaging & semi-implicit, harmonic averaging & explicit, arithmetic averaging & explicit, arithmetic averaging \\ 
		  
		  Pressure-Poisson Solver & preconditioned multigrid, Krylov subspace & not specified & preconditioned multigrid, Conjugate Gradient & preconditioned GMRES, Krylov subspace & preconditioned multigrid, Conjugate Gradient \\

      	  	\bottomrule
	  \end{tabular}}
      \end{table}
  \end{landscape}

%
%
%
%

\pagebreak
\clearpage

\bibliography{paper}
\bibliographystyle{elsarticle-harv}

\end{document}